\documentclass[11pt]{article}
\usepackage{graphicx}
\usepackage{color}
\usepackage{epsfig}
\usepackage{cite}
\usepackage{amsfonts}
\begin{document}
\begin{center}
\title[ \textbf{Thermodynamic Geometry and Hawking Radiation}\\
\vspace{0.5cm} {{\bf  S. Bellucci\footnote{e-mail:
bellucci@lnf.infn.it }$^{,a}$ and B. N. Tiwari\footnote{e-mail:
bntiwari.iitk@gmail.com}$^{,a}$}}
\end{center}
\vspace{0.3cm}
\begin{center}
{$^a$ \it INFN- Laboratori Nazionali di Frascati,\\
 Via E. Fermi 40, 00044, Frascati, Italy}\\
\end{center}
\vspace{1.3cm} This work explores the role of thermodynamic
fluctuations in the two parameter Hawking radiating black hole
configurations. The system is characterized by an ensemble of
arbitrary mass and radiation frequency of the black holes. In the
due course of the Hawking radiations, we find that the intrinsic
geometric description exhibits an intriguing set of exact pair
correction functions and global correlation lengths. We
investigate the nature of the constant amplitude radiation and
find that it's not stable under fluctuations of the mass and
frequency. Subsequently, the consideration of the York model
decreasing amplitude radiation demonstrates that thermodynamic
fluctuations are globally stable in the small frequency region. In
connection with quantum gravity refinements, we take an account of
the logarithmic correction into the constant amplitude and York
amplitude over the Hawking radiation. In both considerations, we
notice that the nature of the possible parametric fluctuations may
precisely be ascertained without any approximation. In the
frequency domain $w \in (0, \infty)$, we observe that both the
local and the global thermodynamic fluctuations of the radiation
energy flux are stable in the s-channel. The intrinsic geometry
exemplifies a definite stability character to the thermodynamic
fluctuations, and up to finitely many topological defects on the
parametric surface, the notion remains almost the same for both
the constant amplitude and the York model. The Gaussian
fluctuations over equilibrium radiation energy flux and
fluctuating horizon configurations accomplish a well-defined,
non-degenerate, curved and regular intrinsic Riemannian manifolds,
for all the physically admissible domains of the radiation parameters.\\
\\
\vspace{2.cm}

\newpage

\begin{Large} \textbf{Contents:} \end{Large}
\begin{enumerate}
\item{Introduction.}
\item{Thermodynamic Geometries.}
\item{Horizon Perturbation:}
\subitem{$3.1$\ \ Constant Amplitude Fluctuations.}
\subitem{$3.2$ \ \ Decreasing Amplitude Fluctuations.}
\item{Quantum Gravity Correction:}
\subitem{$3.1$\ \ Constant Amplitude Fluctuations.}
\subitem{$3.2$ \ \ Decreasing Amplitude Fluctuations.}
\item{Hawking Radiation Energy Flux:}
\subitem{$3.1$\ \ Constant Amplitude Fluctuations.}
\subitem{$3.2$ \ \ Decreasing Amplitude Fluctuations.}
\item{Conclusion and Outlook.}
\end{enumerate}
\section{Introduction}
Since the Hawking's discovery of black hole radiation and quantum
physics \cite{Hawking}, many subsequent aspects of the
semiclassical analysis \cite{semiclassical} have been
investigated. From the viewpoint of the semiclassical theory of
gravity, the present study explores the limiting nature of two
parameter Hawking radiating black holes from the perspective of
fluctuation theory. Importantly, the present anticipation shares
the notion of a very controversial issue, i.e. that of determining
the role of arbitrarily large transplanckian frequencies of vacuum
fluctuations \cite{4,5,6,7,8} and the gravitational back reaction
due to a specific quantum \cite{11}. Besides the `spontaneous'
metric fluctuations, such a discussion takes an account of the
fact that there exist induced metric fluctuations, generated by
the fluctuations of the other quantum fields interacting with the
gravitational one. In the regime where the induced metric
fluctuations dominate, the problem of black hole fluctuations and
back reaction is consistently described by (i) Schwinger-Keldysh
effective stochastic semiclassical theory of gravity \cite{13} and
(ii) Feynman-Vernon influence functional methods \cite{15,16}.
Thus, the semiclassical approximation leads to the stochastic
theory of gravity, extending Einstein equations to the generalized
Einstein-Langevin equations. Such a stochastic stress-energy
tensor describes the most promising theory of the quantum
gravity\cite{18,19} from the phenomenon of the induced metric
fluctuations.

Apart from the notion of the semiclassical gravity
\cite{semiclassical} and the subsequent understanding of black
hole physics \cite{FNB}, the quantum mechanical questions make the
validity doubtful of this semi-classical evolution, which has
received so much attention in the past. Thus, the present paper
sets-up an intrinsic geometric investigation, for analyzing the
horizon properties of a class of two parameter family of Hawking
radiation black hole solutions. An appropriate description is
expected to stem from the background quantum fluctuations. Thus,
in order to determine their effects on the Hawking radiation, one
requires the full theory of the quantum gravity, which is far from
the reach of the present understanding of the subject. It is worth
mentioning further that the study of the black hole fluctuations
problem is technically a complicated issue. However, their
statistical effects connected with the limiting interactions can
be explored from the intrinsic thermodynamic geometric
perspective. Having been motivated from the fluctuating horizon
geometry and Hawking radiation \cite{BFP}, we first consider the
issue how the black hole horizon fluctuations affect the
thermodynamic properties of underlying statistical configuration.
Secondly, we analyze the effects arising from the energy flux
fluctuations of an underlying system, containing an ensemble of
Hawking radiated particles and the Hawking radiating black hole.

In this paper, we take an account of York model \cite{York} and
its possible alternatives arising from the consideration of
fluctuating geometry near the horizon of the black hole. The
configuration of interest is represented by a Vaidya-type metric
with a fluctuating mass. From the viewpoint of the statistical
fluctuation theory, we determine how these fluctuations modify the
horizon configuration, thermodynamic properties, energy flux of
the radiation and what is the asymptotic behavior of the most
dominant s-wave spectrum, originating from the zero point
fluctuations of the background quantum fields, causing the notion
of the generalized uncertainties \cite{bntgup}. Thus, after
neglecting the scattering of the gravitational potential arising
from the 4-dimensional D'Alembertian profile, we may take an
account of the spherically symmetric fluctuations. In order to do
so, let us share a parallel view with a set of important recent
studies, offering the thermodynamic nature of the horizon
properties of diverse (rotating) black holes. This elucidates
interesting aspects of phase transitions, if any, in the
thermodynamic geometric framework and their associated relations
with extremal black hole solutions in the context of $\mathcal N
\geq 2$ compactifications \cite{sfm}. It is worth mentioning
further that the connection of such a geometric formulation to the
thermodynamic fluctuation theory of black holes in general
relativity and in string theory requires several modifications
\cite{rup3}.

Such a geometric formulation has first been applied to $\mathcal
N\geq 2$ supergravity extremal black holes in $D=4$. These
solutions are described as the low energy effective field
theories, arising from the compactifications of Type II string
theories on a compact manifold, \textit{i.e.} $K_3$, Calabi-Yau
and the other associated manifolds \cite{fgk}. Since then, there
have been numerous investigations and several authors have
attempted to understand the possible connections and associated
vacuum phase transitions, if any, which involve some change of the
black hole horizon topology
\cite{cai1,gr-qc/0304015v1,0510139v3,Arcioni}. Intrinsic geometric
modelings involving equilibrium configurations of the extremal and
non-extremal black holes in string theory
\cite{9601029v2,9411187v3,9504147v2,0409148v2,9707203v1,0507014v1,0502157v4,0505122v2}
and $M$-theory
\cite{0209114,0401129,0602015v3,0408106,0408122,SJJS,BBJS,GLS}
possess rich intrinsic geometric structures
\cite{0606084v1,SST,bnt, BNTBull,SAMpaper}. Further interesting
discussions \cite{bntSb, BSBR} are aimed to understand the case of
supersymmetric and non-supersymmetric black holes in various
spacetime dimensions and rotating black string and ring solutions
in the five dimensional spacetimes.

We are motivated from Ruppenier's advocation ``that all the
statistical degrees of freedom of black hole live on the black
hole event horizon''. Thus, the scalar curvature signifies the
average number of correlated Planck areas on the event horizon,
\textit{e.g.}, see for the Kerr-Newman black holes
\cite{RuppeinerPRD78}. Besides several notions analyzed in
condensed matter physics
\cite{RuppeinerRMP,RuppeinerA20,RuppeinerPRL,
RuppeinerA27,RuppeinerA41,RuppeinerPRD78}, we consider specific
Hawking radiating configurations with a finite set of radiation
parameters and analyze possible parametric pair correlations and
their correlation relations. While the Hawking radiation is going
on, the intrinsic geometric computations turn out to be highly
non-trivial, and thus this paper restricts the intrinsic geometric
exploration for the two parameter Hawking radiating black hole
configurations. Following Ruppenier's argument, we take an account
of the fact that the zero scalar curvature indicates certain bits
of information on the event horizon fluctuating independently of
each other, while the diverging scalar curvature signals a phase
transition, indicating highly correlated pixels of information.
Bekenstein has introduced an elegant picture for the quantization
of the area of the event horizon, being defined in terms of Planck
areas \cite{Bekenstein}. Such issues serve as motivations for
considering the quantum correction to the limiting thermodynamic
geometric configuration.

Recently, the thermodynamic geometries of the equilibrium systems
thus anticipated have extensively been explored to investigate the
thermodynamic nature of the limiting configuration of a class of
(rotating) black holes and field theory configurations
\cite{bnt,BNTBull,SAMpaper, BSBR, BNTSBVC}. It is worth mentioning
that there exists an intriguing relationship between the scalar
curvature of the thermodynamic intrinsic Riemannian geometry and
the correlation volume of the corresponding phase-space
configuration, which are both characterized by the parameters of
the radiating black hole. Furthermore, the general coordinate
transformations on the limiting thermodynamic manifolds thus
considered expound to certain microscopic duality relations
associated with the fundamental invariant quantities of the black
hole configuration, \textit{viz.}, mass, charges and chemical
potentials describing the considered Hawking radiation.
Additionally, Ruppeiner has revived the subject with the fact that
the state-space scalar curvature remains proportional to the
correlation volume, which reveals related information residing in
the microscopic models \cite{RuppeinerA20}. Notice further that
the state-space scalar curvature in general signifies a possible
interaction in the underlying radiating configuration. The present
paper is thus intended to analyze both pictures, \textit{viz.},
horizon fluctuations and quantum gravity corrections.

From the perspective of an intrinsic Riemannian geometry, indeed
our study strikingly provides a set of exact covariant
thermodynamic geometric quantities ascribed to a Hawking radiating
black hole. There exists an obvious mechanism on the black hole
side, and it would be interesting to illuminate an associated
quantum gravity notion \cite{ZCZ} for the statistical correlations
to the microstates, or an ensemble of gravity particles arising
from the Hawking radiating systems and vice-versa. In the context
of the Hawking radiation of a neutral black hole, a microscopic
construction follows from the computation of the moments. In the
framework of W-infinity algebra \cite{BonoraCvitan}, the Hawking
radiation arising from a neutral black hole are realized via the
trace anomaly consideration. The finite $W$-algebras, super
algebras and their explicit field realizations have been
considered from the perspective of the $W$-strings, affine
currents and conformal realizations \cite{BSKS1,BSKS2,BSKS3,
BSKS4}. In this concern, it turns out that the present
construction elucidates certain fundamental issues, such as
statistical interactions and the stability of the underlying black
hole configurations, with fluctuating mass and radiation
frequency. Nevertheless, one may arrive to a definite possible
realization of the near equilibrium statistical structures, and
thus the possibility to determine the behavior of the limiting
thermodynamic fluctuations, in terms of the parameters of an
ensemble of limiting equilibrium configurations. Our hope is that
finding statistical mechanical models with similar behavior might
yield further insight into the microscopic properties of the black
hole, and thus leading to a conclusive physical interpretation of
the thermodynamic curvatures and related intrinsic geometric
invariants.

As a matter of fact, a few simple manipulations illustrate that
the relation of a non-zero scalar curvature with an underlying
interacting statistical system remains valid even for higher
dimensional intrinsic Riemannian manifolds, and the connection of
a divergent scalar curvature with phase transitions may
accordingly be divulged from the Hessian matrix of the considered
energy flux/ fluctuating horizon area. It is worth mentioning that
our analysis takes an intriguing account of the scales that are
larger than the correlation length and considers that only a few
microstates do not dominate the whole macroscopic equilibrium
intrinsic quantities. Specifically, we share the interpretation
that the underlying horizon perturbations and energy fluxes
include contributions from a large number of subensembles of the
fluctuating horizon black holes. In this sense, such a
consideration takes an account of the fact that the saddle point
approximation leads to an appropriate physics of the fluctuating
horizon. Thus, the present investigation characterizes the
covariant intrinsic geometric description for understanding the
quantum statistical physics \cite{ZCZ} of a Hawking radiating
black hole.

With this motivation, our geometric formulations tacitly involves
unified statistical basis, in terms of the chosen ensemble. In
particular, we have outlined some space-time implications of the
thermodynamic geometry for the two parameter Hawking radiating
configurations. The underlying motivations arising from the
microscopic quantum gravity configuration turn out to have a
bootstrapping significance. Although the analysis has only been
considered in the limit of small statistical fluctuations, however
the underlying correlation length takes an intriguing account upon
the quartic corrections of the energy flux, or the horizon area of
the fluctuating black hole. Herewith, we focus our attention on
the interpretation that the underlying energy includes
contributions from a large number of excited particles, described
in the framework of quantum field theories on curved space-time
geometry. Thus, our description of the geometric thermodynamics
extends itself to all possible configurations. Some of these
issues have been anticipated in the final section of the paper.

The present paper is organized as follows. In the first section,
we have presented the motivations to study the intrinsic
Riemannian surfaces obtained from the Gaussian fluctuations of a
Hawking radiating black hole configuration. In section $2$, we
briefly explain what are the intrinsic thermodynamic geometries
based on an ensemble of equilibrium statistical basis
characterizing a two parameter Hawking radiating black hole. In
section $3$, we obtain the two parameter canonical configurations
with a constant amplitude and with decreasing amplitude of
background fluctuations. In section $4$, we investigate the
corrections to the above considered thermodynamic geometric
configurations, arising from the quantum gravity fluctuations. In
section $5$, we focus our attention on the thermodynamic geometry
of the aforementioned configurations, with a fluctuating radiation
energy flux. In all the above cases, we have explained that such
an intrinsic geometric configuration, obtained from an effective
energy flux or fluctuating horizon area of the black holes,
results to be well-defined and pertains to an interacting
statistical system. Finally, section $6$ contains a set of
concluding issues and a possible perspective discussion of the
Hawking radiating black holes and their thermodynamic geometries.
The general implications thus obtained may divulge the intrinsic
geometry of both the chemical and equilibrium microscopic
acquisitions. These issues are the matter of a future
investigation.
\section{Thermodynamic Geometries}

The present section presents a brief review of the essential
features of thermodynamic geometries from the perspective of the
application to the two parameter Hawking radiating configurations.
From the viewpoint of the Hawking radiation, the present analysis
considers two parameter radiating black holes. Subsequently, we
divulge the most probable fluctuations around a chosen equilibrium
black hole configuration. Thereby, we provide provide arguments to
understand the most puzzling black hole, \textit{viz.}, a
radiating Schwarzschild black hole.

The underlying intrinsic configuration is parameterized by the
mass $M$ and frequency $w$ of the radiation. The thermodynamic
fluctuations around a minimum energy flux $F(M,w)$ (or maximum
horizon area $A(M,w)$ are described as an intrinsic configuration.
The concerned local pair correlations are characterized by the
metric tensor of the associated Wienhold(/ Ruppenier) surface.
Consequently, both of the above two dimensional intrinsic
Riemannian geometry are conformally the same. Thus, up to a
constant factor,
the components of the thermodynamic metric tensor are given by \\
\begin{scriptsize}
\begin{eqnarray} \label{metric}
g_{M M}&=& \frac{\partial^2 A}{\partial M^2} \nonumber  \\
g_{M w}&=& \frac{\partial^2 A}{{\partial M}{\partial  w}}\nonumber  \\
g_{w w}&=& \frac{\partial^2 A}{\partial
w^2}
\end{eqnarray}
\end{scriptsize}
In this case,  it follows that the determinant of the metric tensor is
\begin{scriptsize}
\begin{eqnarray}
\Vert g(M,w) \Vert &= &A_{MM}A_{w w}- A_{M w}^2
\end{eqnarray}
\end{scriptsize}
Explicitly, we can calculate the $\Gamma_{ijk}$, $R_{ijkl}$, $R_{ij}$
and $ R $ for the above two dimensional thermodynamic geometry $(M_2,g)$.
One may easily inspect that the scalar curvature is given by
\begin{scriptsize}
\begin{eqnarray}
R(M, w)&=& -\frac{1}{2} (A_{MM}A_{ ww}- A_{M  w}^2)^{-2} (A_{ w w}A_{MMM}A_{M  w w}\nonumber
\\ &&+ A_{M  w}A_{MM w}A_{M w w}+A_{MM}A_{MM w}A_{ w w w}\nonumber  \\
&&- A_{M w}A_{MMM}A_{w w w}- A_{MM}A_{M w w}^2 -A_{ w w}A_{MM w}^2 )
\end{eqnarray}
\end{scriptsize}
Furthermore, the relation between the intrinsic scalar curvature
and the Riemann covariant curvature tensor for any two dimensional
intrinsic geometry is given (see for details \cite{bnt}) by
\begin{scriptsize}
\begin{eqnarray} \label{cur}
R(M, w)=\frac{2}{\Vert g(M, w) \Vert}R_{M wM w}(M, w)
\end{eqnarray}
\end{scriptsize}
The relation in Eqn.\ref{cur} is standard for an arbitrary
intrinsic surface $(M_2(R),g)$. Correspondingly, the Legendre
transformed version of the energy fluctuation manifold is known as
a state-space manifold with a fluctuating entropy configuration.
Subsequently, we investigate the nature of the thermodynamic thus
defined, as an intrinsic Riemannian manifold obtained from the
horizon perturbations, with the leading order contributions being
considered in an underlying ensemble. Such an analysis continues,
even in the context of chemical reactions or in any closed system,
and it strongly suggests that a non-zero scalar curvature might
provide useful information regarding the range of underlying
phase-space correlations between various components of an
underlying Hawking radiating black hole configuration.
Furthermore, our results are expected to be in a close connection
with the microscopic implications, if any, arising from the
underlying field theory theories. Thus, these results yield an
anticipation to offer physically definite explanations of the
Hawking radiating configurations.
\section{Horizon Perturbation}
With the general introduction of the thermodynamic geometric nature of Hawking radiating
black holes, we shall now define the negative (positive) Hessian function of the horizon
area (energy flux) and thereby proceed to investigate the thermodynamic behavior of the
two parameter Hawking radiating black holes and their thermo-geometric structures.
\subsection{Constant Amplitude Fluctuations}
In order to begin the intrinsic geometric analysis of
thermodynamic fluctuations, we consider the correlation in the
configuration arising from the fluctuations of the radiation
parameters. Thus, we introduce an ensemble of black holes
fluctuating over the limiting Gaussian configuration. In this
analysis, we consider that the black hole can have non-zero
fluctuations due to the vibrations of frequencies, residual
factors in vacuum, and possibly other quantum properties. This
follows from the fact that we do not restrict ourselves to the
specific domains of radiation and particle productions.

Consequently, we allow for an ensemble of limiting horizon
configurations with finite fluctuations in an arbitrary non linear
domain of the parameters and thereby analyze the nature of a class
of generic limiting thermodynamic evolutions. We further take an
account of the variable radiation frequency defining the speed
response tuning parameter of the particles. Subsequently, the
stability of the system can be analyzed in the entire mass and
frequency domains of the radiation. This is observed, when there
are back reactions in the black hole. In this framework, we find
that the area of the fluctuating horizon, as the function of mass
and radiation frequency, takes an intriguing expression
\begin{scriptsize}
\begin{eqnarray} \
A(M,w) := 16 \pi M^2 (1+ \frac{u^2}{(2+16 M^2 w^2)})
\end{eqnarray}
\end{scriptsize}
The correlations are described by the Hessian matrix of the
horizon area, defined with a set of desired corrections over a
chosen black hole with adjustable frequency band. Following
Eqn.(\ref{metric}), the components of the metric tensor, defined
as the Hessian function of the horizon area $Hess(A(M,w))$, reduce
to the following expressions
\begin{scriptsize}
\begin{eqnarray} \label{CCmetric}
g_{MM} &=& -16 \pi \frac{(1024 M^6 w^6 +384 M^4 w^4 +28 M^2 w^2(2- u^2)+2 +u^2)}{(1+8 M^2 w^2)^3} \nonumber \\
g_{Mw} &=& 512 \pi \frac{M^3 u^2 w}{(1+8 M^2 w^2)^3} \nonumber \\
g_{ww} &=& -128 \pi \frac{M^4 u^2 (24 M^2 w^2-1)}{(1+8 M^2 w^2)^3}
\end{eqnarray}
\end{scriptsize}
In this framework, we observe that the geometric nature of
parametric pair correlations divulges the notion of a fluctuating
Hawking radiating black hole. The fluctuating black hole may thus
be easily determined in terms of the intrinsic parameters of the
underlying configurations.
\begin{figure} \label{Gcc}
\hspace*{0.5cm}
\includegraphics[width=8.0cm,angle=-90]{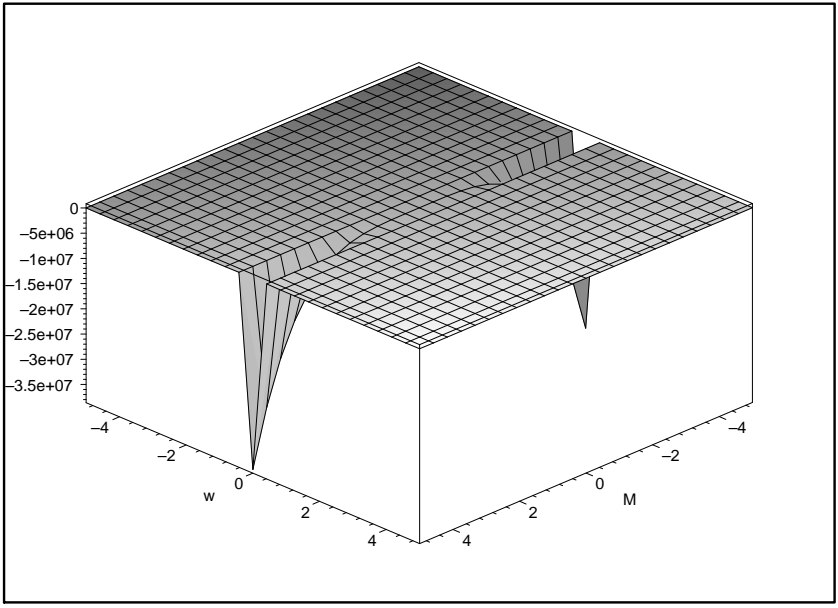}
\caption{The determinant of the metric tensor plotted as the function of the radiation
frequency $w$ and mass $M$, describing the fluctuations in the constant amplitude Hawking
radiating black hole.}
\vspace*{0.5cm}
\end{figure}
It is worth mentioning that the two parameter black hole turns out
to be well-behaved for the generic values of the mass and
radiation frequency. Over the domain of radiation parameters $\{
M, w \}$, we notice that the Gaussian correlations form stable
correlations, if the determinant of the metric tensor
\begin{scriptsize}
\begin{eqnarray} \label{CCdet}
\Vert g(M,w) \Vert = 2048 \pi^2 \frac{M^4 u^2}{(1+8 M^2 w^2)^5} g_c(M,w)
\end{eqnarray}
\end{scriptsize}
remains a positive function on the parametric surface $(M_2(R),g)$.
The stability of the radiation is thus defined by the positivity of
\begin{scriptsize}
\begin{eqnarray} \label{g_c(M,w)}
g_c(M,w):= 3072 M^6 w^6+ 640 M^4 w^4+ 8w^2(2-9 u^2) M^2- 2- u^2
\end{eqnarray}
\end{scriptsize}
Thus, the cubic equation $g_c(M,w)=0$ determines the thermodynamic
stability structure of the constant amplitude Hawking radiating
black holes. This follows simply from the fact that the
Eqn.(\ref{g_c(M,w)}) is a cubic equation in $M^2$, and thus the
positivity of $g_c(M,w)$ could be realized in a specific domain of
the radiation amplitude. It is not difficult to compute an exact
expression for the scalar curvature describing the global
parametric intrinsic correlations. By defining a radiation
function $R_c(M,w)$, we find that the most general scalar
curvature can explicitly be presented as
\begin{scriptsize}
\begin{eqnarray} \label{CCcur}
R(M,w) = -\frac{1}{2\pi M^2} \frac{R_c(M,w)}{g_c(M,w)^2}
\end{eqnarray}
\end{scriptsize}
where the global interaction of the radiation is defined by the function
\begin{scriptsize}
\begin{eqnarray}
R_c(M,w)&:=&589824 M^{10} w^{10}+204800 M^8 w^8+(22528 w^6-4608 w^6 u^2) M^6 \nonumber \\&&
+(-576 u^2 w^4+768 w^4)M^4+(8 u^2 w^2+16 w^2) M+2+u^2
\end{eqnarray}
\end{scriptsize}
Thus, the present intrinsic geometric analysis anticipates that
the constant amplitude Hawking radiating black hole is always an
interacting statistical system over the surface of the fluctuating
mass and radiation frequency, except on the real roots of the
equation $R_c(M,w)=0$. These points correspond to a
non-equilibrium statistical system, and their numerical values may
precisely be easily obtained from the roots of the following
degree five equation
\begin{scriptsize}
\begin{eqnarray}
589824 w^{10} p^{5}+204800 w^8 p^4+(22528 w^6-4608 w^6 u^2) p^6
+(-576 u^2 w^4+768 w^4)p^2+(8 u^2 w^2+16 w^2)p+2+u^2=0,
\end{eqnarray}
\end{scriptsize}
where $p:=M^2$. Notice further that the limit $M=0$ is unstable
for the Hawking radiating black hole. It is expected that an
extremely small (large) mass black hole goes beyond the validity
of the saddle point approximation of the Euclidean path integral.
Thus, it may be anticipated that the higher order quantum gravity
corrections would dominate the physical observations, in such
extreme domains. Interestingly, the quantum notion may further be
made obvious from the very basic addition of the background
space-time fluctuations to the spectrum of the theory.
Furthermore, such an initiation would be the matter of the next
section of the present investigation. From the thermodynamic
perspective, we would thereby explore what could statistically
happen to a black hole, while the Hawking radiations are in the
verge of an extinction.

In the present case the fluctuating horizon geometries, the
thermodynamic stability hypothesis follows directly from the sign
of the determinant of the metric tensor. The corresponding horizon
analysis shows that the amplitude becomes a purely imaginary
number \textit{viz.}, $u= \sqrt{2} i$ in the limit of $M=0$ or
$w=0$. What follows further that we specialize ourselves to the
limiting value of the radiation frequency, and subsequently, we
analyze the stability for $w=0$ corresponding to the limiting
equilibrium statistical basis. Physically, the limiting
equilibrium horizon is realized with no Hawking radiation. Thus,
for the black hole thus achieved, the zero frequency configuration
has the following limiting local thermodynamic correlations
\begin{scriptsize}
\begin{eqnarray}
g_{MM} &=& -16 \pi(2+u^2) \nonumber \\
g_{Mw} &=& 0 \nonumber \\
g_{ww} &=& 128 \pi M^4u^2
\end{eqnarray}
\end{scriptsize}
Under such a limiting specification of the parameters, the global
quantities, \textit{viz.}, the determinant of the metric tensor
and Ricci scalar are reduced to the following limiting values
\begin{scriptsize}
\begin{eqnarray}
\Vert g(M,w=0) \Vert &=& -2048 \pi^2 M^4 u^2 (2+u^2) \nonumber \\
R(M,w=0) &=& -\frac{1}{2 \pi M^2 (2+u^2)}
\end{eqnarray}
\end{scriptsize}
\begin{figure} \label{Rcc}
\hspace*{0.5cm}
\includegraphics[width=8.0cm,angle=-90]{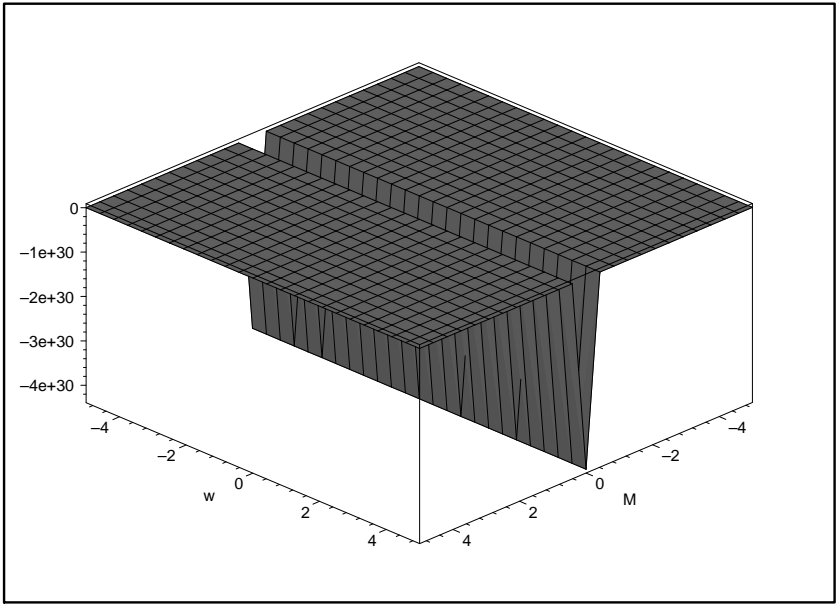}
\caption{The curvature scalar plotted as the function of the radiation frequency
$w$ and mass $M$, describing the fluctuations in the constant amplitude Hawking
radiating black hole.}
\vspace*{0.5cm}
\end{figure}
Significantly, the behavior of the determinant of the metric
tensor shows that such limiting black holes become unstable for
all possible physical values of the mass. For $u=1$, the pictorial
nature of the determinant of the metric tensor is depicted in
Fig.[1]. We observe that the system is stable in the zero mass
limit, but acquires a large negative fluctuation of order
$10^{7}$. As expected on physical grounds, it is worth mentioning
further that the limiting zero mass black hole becomes highly
self-contracting. As a matter of fact, Fig.[2] shows that the
limit of vanishing mass yields a negative strength of the
thermodynamic interaction in the order of $10^{30}$. In this
process. we observe further that the system acquires a triangular
wall of instability, in the limit of $\vert M \vert \rightarrow
0$.

Physically, if one shares the notion that the black hole has a
remnant, then the above picture changes slightly, and we find that
the small mass black holes are unstable with a high degree of
self-interaction. A refined analysis is the matter of the next
section, where we shall take an account of the quantum gravity
corrections. Specifically, it is important to mention that the
correlation length of the underlying nearly equilibrium system is
globally characterized by the scalar curvature of $(M_2,g)$. The
present investigation shows that a typical Hawking radiating black
hole is globally correlated over all possible Gaussian
fluctuations of the frequency and a non-zero mass.
\subsection{Decreasing Amplitude Fluctuations}
In this section, we study the thermodynamic geometry arising from
the horizon area of a Hawking radiating black hole. It is well
known that the extremal black holes do not Hawking radiate, since
the Hawking temperature is proportional to the difference of the
inner and outer horizon radius of the hole configuration.
Therefore, we focus our attention of the non-extremal
configuration and analyze the thermodynamic properties of such
black holes with a large mass $M \gg m$, where $m$ is the Planck
mass. Following York \cite{York} we assume the dimensionless
amplitude
\begin{scriptsize}
\begin{eqnarray} \label{decamp}
u = a (\frac{m}{M}),
\end{eqnarray}
\end{scriptsize}
where $a$ is a pure number. In particular, the assumption $a \ll 1
$ means that the black hole is far away from the Planck size black
holes. In order to get a more realistic result, it is assumed that
one should average over the entire spectrum of the space-time
metric fluctuations. This is realized with fluctuating the
parameters $\{M,w\}$ of the black hole space-time. In this
section, $A(M,w)$ denotes the horizon entropy of the black hole,
since up to a normalization both quantities are the same for the
present discussion. With the consideration of the York model
decreasing amplitude Eqn.\ref{decamp}, we find that the horizon
entropy is given by the following expression
\begin{scriptsize}
\begin{eqnarray}
A(M,w) := 4 \pi (\frac{M}{m})^2-4 \pi \frac{a^2}{(1+16 M^2 w^2)}
\end{eqnarray}
\end{scriptsize}
While the Hawking radiation is going on, there exist non-trivial
intrinsic correlations. The components of the thermodynamic metric
tensor, defining fluctuation among the parameters, are given by
\begin{scriptsize}
\begin{eqnarray} \label{CDmetric}
g_{MM} &=&  8 \pi \frac{(-4096 M^6 w^6-768 M^4 w^4+(768 a^2 w^2 m^2-48)w^2 M^2-1-16 a^2 w^2 m^2)}{m^2(1+16 M^2 w^2)^3} \nonumber \\
g_{Mw} &=& 256 \pi a^2 M w \frac{(16 M^2 w^2-1)}{(1+16 M^2 w^2)^3} \nonumber \\
g_{ww} &=&  128 \pi a^2 M^2 \frac{(48 M^2 w^2-1)}{(1+16 M^2 w^2)^3}
\end{eqnarray}
\end{scriptsize}
For such Hawking radiating black holes, we observe that the
principle components of the metric tensor $ \{g_{MM}, g_{ww}\}$
signifying self pair correlations, are positive definite functions
over a range of the parameters. Physically, this signifies a set
of heat capacities against the intrinsic interactions, arising
from a fictitious potential flowing on the surface $(M_2(R),g)$ of
an ensemble of Hawking radiated particles. Moreover, it is evident
that the  determinant of the metric tensor reduces to the
following simple expression
\begin{scriptsize}
\begin{eqnarray} \label{CDdet}
\Vert g(M,w) \Vert = 1024 \pi^2 \frac{a^2 M^2}{m^2 (1+16 M^2 w^2)^5} g_d(M,w)
\end{eqnarray}
\end{scriptsize}
where stability of the decreasing amplitude radiation is determined by the positivity of the function
\begin{scriptsize}
\begin{eqnarray}
g_d(M,w):= -12288 M^6 w^6-1280 M^4 w^4+ 16(80 a^2 w^2 m^2-1) M^2-48 a^2 w^2 m^2+1
\end{eqnarray}
\end{scriptsize}
\begin{figure}
\hspace*{0.5cm}
\includegraphics[width=8.0cm,angle=-90]{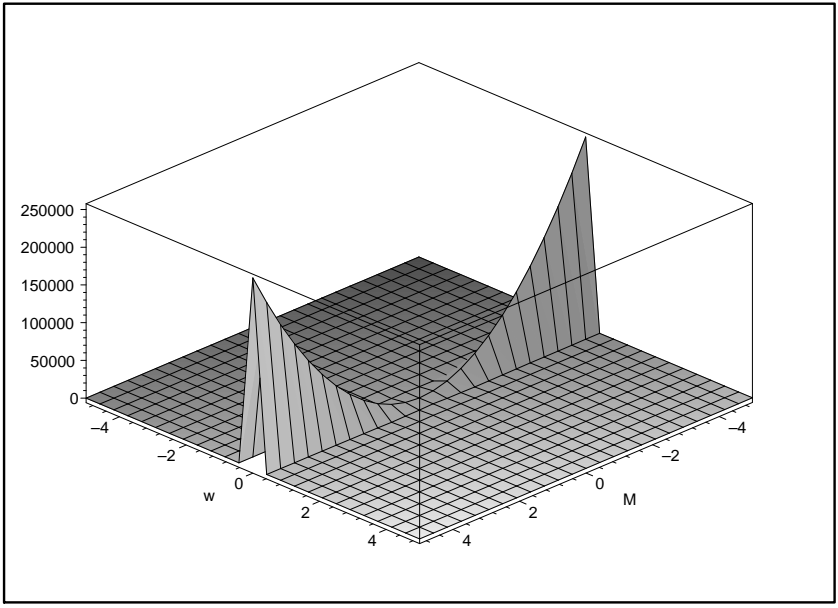}
\caption{The determinant of the metric tensor plotted as the function of the radiation
frequency $w$ and mass $M$, describing the fluctuations in the decreasing amplitude Hawking
radiating black hole.}
\vspace*{0.5cm}
\end{figure}
As in the case of a constant amplitude Hawking radiating black
hole, we observe in the present case that the stability structure
is again determined by a cubic equation. This follows from the
fact that the sign is governed by the variable $M^2$. Furthermore,
the configuration in which the cubic equation $g_d(M^2,w)=0$ has a
real positive value is stable. Thus, it is evident that the global
stability of the fluctuating Hawking radiating decreasing
amplitude configuration is determined by the Ricci scalar
curvature of the parametric surface. For the above metric tensor
Eqn.(\ref{CDmetric}), we may easily obtain a compact formula for
the scalar curvature. Exclusively, we find in the present case
that the intrinsic geometric analysis assigns a compact expression
to the invariant quantity. For an arbitrary mass and frequency of
the Hawking radiation, a straightforward computation shows that
the most general scalar curvature is given by the ratio of the two
polynomials of the radiation parameters
\begin{scriptsize}
\begin{eqnarray} \label{CDcur}
R(M,w) = \frac{m^2}{4\pi M^2} \frac{R_d(M,w)}{g_d(M,w)^2}
\end{eqnarray}
\end{scriptsize}
where the global interaction of the decreasing amplitude fluctuating radiation is defined by the function
\begin{scriptsize}
\begin{eqnarray}
R_d(M,w)&:=& 150994944 w^{12} M^{12}-2818048 w^8 M^8+(1572864 a^2 w^2 m^2-163840) w^6 M^6 \nonumber \\&&
+(65536 a^2 w^2 m^2+768) w^4 M^4+(-2048 a^2 w^2 m^2+128) w^2 M^2-1
\end{eqnarray}
\end{scriptsize}
\begin{figure}
\hspace*{0.5cm}
\includegraphics[width=8.0cm,angle=-90]{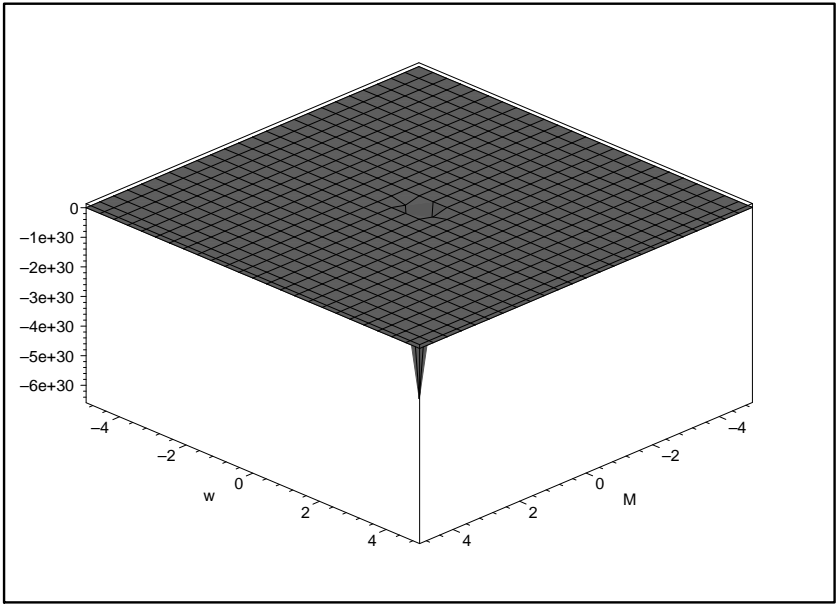}
\caption{The curvature scalar plotted as the function of the radiation frequency
$w$ and mass $M$, describing the fluctuations in the decreasing amplitude Hawking
radiating black hole.}
\vspace*{0.5cm}
\end{figure}
The notion of the fluctuations is easily determined in terms of
the intrinsic parameters of the radiating black hole. It is worth
mentioning that the thermodynamic equilibrium of the Hawking
radiation could be anticipated to be reached in the limit of zero
Hawking frequency, \textit{viz.}, $w= 0$. In this limit, it
follows that the underlying local thermodynamic pair correlations
of the Hawking radiating black hole with the present type
amplitude reduce to the following limiting values
\begin{scriptsize}
\begin{eqnarray}
g_{MM} &=& -8 \frac{\pi}{m^2} \nonumber \\
g_{Mw} &=& 0 \nonumber \\
g_{ww} &=& -128 \pi a^2 M^2
\end{eqnarray}
\end{scriptsize}
As in the previous section, the global quantities of the
fluctuating parametric surface of a decaying amplitude Hawking
radiating black hole, \textit{viz.}, the determinant of the metric
tensor and the Ricci scalar curvature, respectively reduce to the
following limiting values
\begin{scriptsize}
\begin{eqnarray}
\Vert g(M,w=0) \Vert &=& 1024 \pi^2 a^2 (\frac{M}{m})^2 \nonumber \\
R(M,w=0) &=&-\frac{1}{4 \pi} (\frac{m}{M})^2
\end{eqnarray}
\end{scriptsize}
Herewith, the intrinsic geometric framework makes the nature of
parametric pair correlations more clearly observable. In the limit
of $w=0$, we find that the determinant of the metric tensor is
positive, signifying stable limiting thermodynamic fluctuations
and thus a favorable condition for the stability of the radiating
black hole. Similarly, the negativity of the scalar curvature
indicates that the thermodynamic correlations are attractive on
$(M_2(R),g)$. Notice further that such a viewpoint suggests that
the Planck size black holes would be self-attractive with a
constant correlation length.

Graphically, the behavior of the determinant of the metric tensor
shows that such limiting black holes become unstable in the limit
of small frequency and large mass. For $a=1, m=1$, the pictorial
nature of the determinant of the metric tensor is depicted in the
Fig.[3]. Herewith, we notice that the system is stable in the zero
mass limit, but acquires a large positive fluctuation of order
$10^{5}$. From the Fig.[3,4], we observe for $a=1, m=1$ that the
nature of the decaying amplitude Hawking radiating black hole is
precisely opposite to the constant amplitude Hawking radiating
black hole. As per the anticipation of the present consideration,
the Fig.[4] shows that the limiting zero mass and zero frequency
black hole becomes highly self-contracting. It is worth mentioning
further that the limit of vanishing mass and radiation frequency
makes the negative strength of the thermodynamic global
interaction to be of order $10^{30}$. In this process. we observe
from Fig.[4] that the system acquires a large conical wall of
thermodynamic instability, in the limit of $\vert M \vert
\rightarrow 0$. As a matter of fact, the existence of the black
hole remnant would make the Fig.[4] to be capped-off. Thus, the
underlying thermodynamic singularity would turn out to be a large
finite asymmetric cylinder.
\section{Quantum Gravity Correction}
In the present section, we explore the statistical nature of an
ensemble of generic particles generated from the Hawking radiation
of a massive black hole. As in the previous section, we restrict
our attention to the constant amplitude and decreasing amplitude
radiation black holes. The main goal of the present section is to
describe the thermodynamic behavior of the quantum gravity
fluctuations over the constant amplitude and decreasing amplitude
radiating configurations.
\subsection{Constant Amplitude Fluctuations}
To consider the most general case, we chose the horizon area as
the function of the mass and fluctuation frequency, along with the
incorporation of the non-perturbative quantum corrections to the
horizon area of the black hole \cite{ZCZ}. In this case, the bare
horizon area is given by
\begin{scriptsize}
\begin{eqnarray}
B(M,w) := 16 \pi M^2 (1+\frac{u^2}{(2+16 M^2 w^2)})
\end{eqnarray}
\end{scriptsize}
The quantum corrected horizon area \cite{ZCZ} is expressed as
\begin{scriptsize}
\begin{eqnarray} \label{cqg}
A(M,w) := 4 \pi (\frac{M}{L})^2 (1+\frac{u^2}{(2+16 M^2 w^2)})+a \ln(\frac{16 \pi M^2}{L^2} (1+\frac{u^2}{(2+16 M^2 w^2)})
\end{eqnarray}
\end{scriptsize}
Let us restrict ourselves to a situation where the number of
parameters remains unchanged under the quantum dynamics, producing
an ensemble of particles. When the mass and radiation frequency is
allowed to fluctuate, we can exploit the definition of the Hessian
function $Hess(A(M,w))$ for the Eqn.(\ref{cqg}). For the quantum
gravity corrected Hawking radiating black hole, we see that there
is the following set of parametric pair correlations
\begin{scriptsize}
\begin{eqnarray} \label{QCmetric}
g_{MM} &=& \frac{2}{M^2L^2}
\frac{(g^{cMM}_{12}M^{12}+ g^{cMM}_{10}M^{10}+ g^{cMM}_{8}M^{8}+ g^{cMM}_{6}M^{6}+ g^{cMM}_{4}M^{4}+ g^{cMM}_{2}M^{2}+g^{cMM}_{0})}
{(2+16 M^2 w^2+u^2)^2(1+8 M^2 w^2)^3} \nonumber \\
g_{Mw} &=& -\frac{32Mu^2w}{L^2}
\frac{(g^{cMw}_{6}+M^6+ g^{cMw}_{4}M^4+ g^{cMw}_{2} M^2+ g^{cMw}_{0})}{(2+16 M^2 w^2+u^2)^2(1+8 M^2 w^2)^3} \nonumber \\
g_{ww} &=& -\frac{16M^2u^2}{L^2}
\frac{(g^{cww}_{8}M^8+ g^{cww}_{6}M^6+g^{cww}_{4}M^4+ g^{cww}_{2}M^2+g^{cww}_{0})}{(2+16 M^2 w^2+u^2)^2(1+8 M^2 w^2)^3}
\end{eqnarray}
\end{scriptsize}
where the functions $\{g^{cab}_{2i} \vert \ a,b \in \{ M,w \}; 0
\le i \le 6 \}$ are defined in the Appendix (A1). Subsequently, we
see that the fluctuations of the radiating massive black hole
systems comply with the physically expected conclusions. In
particular, the self-pair correlations, defining a set of heat
capacities, remain positive quantities for a domain of mass and
frequency. A straightforward computation further demonstrates the
over-all nature of the parametric fluctuations. A radiating black
hole is stable under the set of Gaussian fluctuations, if the
associated principle minors $\{g_{MM}, g \}$ remain positive
functions on $(M_2,g)$. Subsequently, an explicit computation
shows that the stability constraint on the surface is given by the
positivity of the determinant of the metric tensor
\begin{scriptsize}
\begin{eqnarray} \label{QCdet}
\Vert g(M,w) \Vert = -\frac{32 u^2 }{L^4} \frac{ \tilde{g}(M,w)}{(2+16 M^2 w^2+u^2)^3(1+8 M^2 w^2)^5}
\end{eqnarray}
\end{scriptsize}
\begin{figure}
\hspace*{0.5cm}
\includegraphics[width=8.0cm,angle=-90]{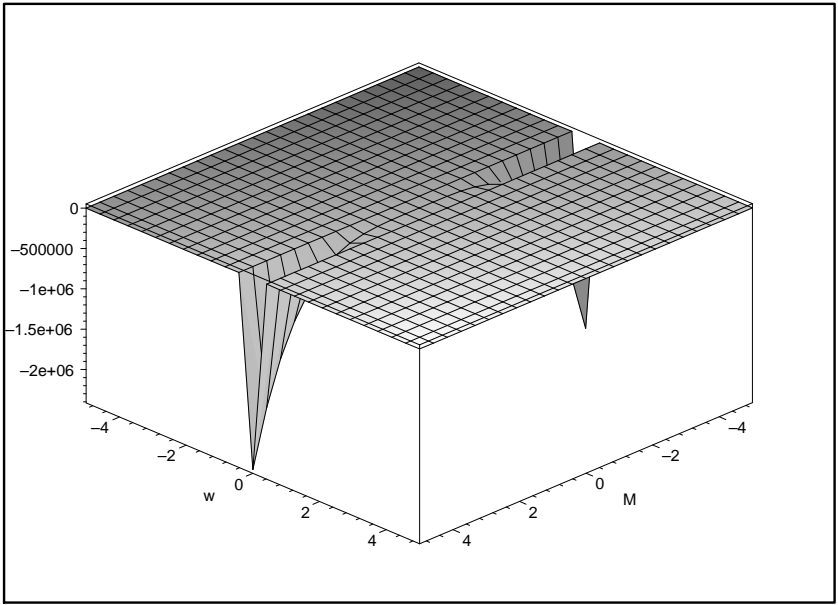}
\caption{The determinant of the metric tensor plotted as the function of the
radiation frequency $w$ and mass $M$, describing the fluctuations in the constant
amplitude Hawking radiating black hole with the quantum gravity corrections.}
\vspace*{0.5cm}
\end{figure}
Notice that the co-ordinate charts on $(M_2,g)$ are described by the parameters $\{M,w\}$ of the radiation.
In the above Eqn.(\ref{QCdet}), the determinant of the metric tensor can be positive, if the function
\begin{scriptsize}
\begin{eqnarray}
\tilde{g}(M,w):= \sum_{n=0}^8 g^{c}_{2n}M^{2n}
\end{eqnarray}
\end{scriptsize}
takes a positive value on the intrinsic thermodynamic surface. To
be specific, we have defined the functions $\{g^{c}_{2n} \vert \ 0
\le n \le 8 \}$ in the Appendix (A2). Notice further that it is
not difficult to compute an exact expression for the scalar
curvature describing the global parametric intrinsic correlations.
By defining a set of functions  $\{r^{c}_{2n} \}$ depending on the
frequency of the Hawking radiation, it turns out that the most
general scalar curvature can be expressed as
\begin{scriptsize}
\begin{eqnarray} \label{QCcur}
R(M,w) = (\frac{Lm}{M})^2
\frac{\sum_{n=0}^{14} r^{c}_{2n} M^{2n}}{(\tilde{g}(M,w))^2}
\end{eqnarray}
\end{scriptsize}
where the radiation functions $\{r^{c}_{2n} \vert \ 0\le n \le 14
\}$ appearing in the numerator of the scalar curvature have been
relegated to the Appendix (A3). Consequently, we notice that
$\{r^{c}_{2n} \}$ can solely be expressed as functions of the
radiating frequency of the black hole.
\begin{figure}
\hspace*{0.5cm}
\includegraphics[width=8.0cm,angle=-90]{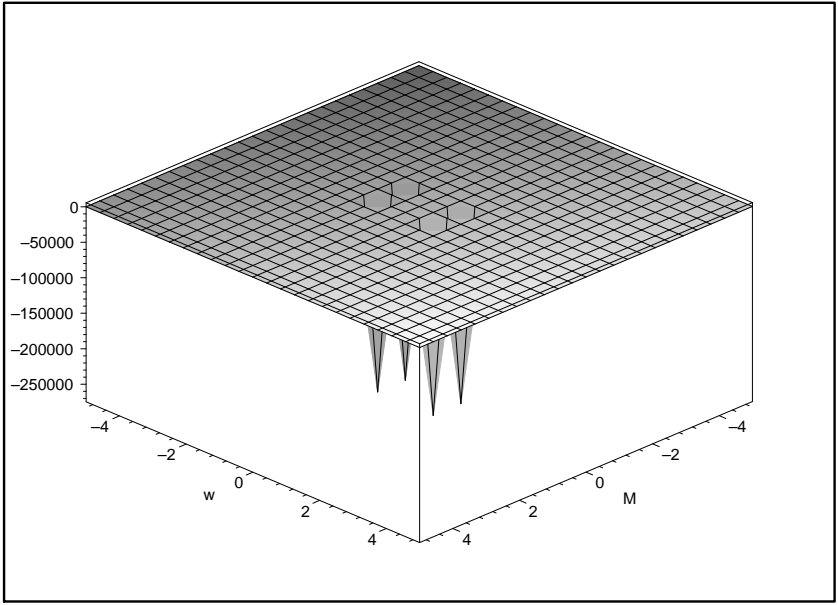}
\caption{The scalar curvature plotted as the function of the radiation frequency $w$ and
mass $M$, describing the fluctuations in the constant amplitude Hawking radiating black
hole with the quantum gravity corrections.}
\vspace*{0.5cm}
\end{figure}
In the limit $w=0$, the local pair correlations reduce to the
following values
\begin{scriptsize}
\begin{eqnarray}
g_{MM} &=& \frac{2}{M^2L^2} (\frac{(-2 \pi u^6-12 \pi u^4-24 \pi u^2-16 \pi) M^2+4 a L^2+a L^2 u^4+4 a L^2 u^2}{(2+u^2)^2} ) \nonumber \\
g_{Mw} &=& 0 \nonumber \\
g_{ww} &=& \frac{16M^2 u^2}{L^2} (\frac{(2 \pi u^4+8 \pi+8 \pi u^2) M^2+a L^2 u^2+2 a L^2}{(2+u^2)^2})
\end{eqnarray}
\end{scriptsize}
The limit $w=0$ reduces the determinant of the metric tensor to the following quantity
\begin{scriptsize}
\begin{eqnarray}
\Vert g(M,w=0) \Vert = -\frac{32 u^2}{L^4} \frac{d^{c}(M,w=0)}{(2+u^2)^3}
\end{eqnarray}
\end{scriptsize}
where the numerator of the limiting determinant of the metric tensor is given by the polynomial
\begin{scriptsize}
\begin{eqnarray}
d^{c}(M,w=0)&:=&(64 \pi^2+32 \pi^2 u^6+128 \pi^2 u^2+4 \pi^2 u^8+96 \pi^2 u^4) M^4\nonumber \\ &&
-4 a^2 L^4-a^2 L^4 u^4-4 a^2 L^4 u^2
\end{eqnarray}
\end{scriptsize}
Notice further that the limiting small frequency configuration is stable over the Gaussian fluctuations,
if the mass of the Hawking radiating black hole with small background fluctuations satisfies
\begin{scriptsize}
\begin{eqnarray}
| M | < (\frac{4 a^2 L^4+a^2 L^4 u^4+4 a^2 L^4 u^2}{64 \pi^2+32 \pi^2 u^6+128 \pi^2 u^2+4 \pi^2 u^8+96 \pi^2 u^4})^{1/2}
\end{eqnarray}
\end{scriptsize}
This follows from the analysis of the algebraic equation $\alpha
M^4- \gamma=0$, which has a real positive root in the domain $M<
\sqrt{\gamma / \alpha}$ for $\alpha, \gamma \in R$. The global
nature of phase transitions can be thus explored over the range of
parameters describing the Hawking radiation of interest. For the
specific Hawking radiations corresponding to the limiting
frequency $w=0$, the limiting intrinsic scalar curvature
simplifies to the following ratio of the two polynomials
\begin{scriptsize}
\begin{eqnarray}
R(M,w=0) = -2 L^2  \frac{n^{c}(M,w=0)}{d^{c}(M,w=0)^2}
\end{eqnarray}
\end{scriptsize}
where the polynomial $n^{c}(M,w=0)$ in the numerator is given by
\begin{scriptsize}
\begin{eqnarray}
n^{c}(M,w=0)&:=& (224 \pi^3 u^12+2048 \pi^3+16 \pi^3 u^14+4480 \pi^3 u^8  \nonumber \\ &&
+10752 \pi^3 u^4+7168 u^2 \pi^3+1344 \pi^3 u^10+8960 \pi^3 u^6) M^6 \nonumber \\ &&
+(-240 a L^2 \pi^2 u^8-48 a L^2 \pi^2 u^10-960 a L^2 \pi^2 u^4 \nonumber \\ &&
-768 a L^2 \pi^2 u^2-640 a L^2 \pi^2 u^6-4 a L^2 \pi^2 u^12-256 a \pi^2 L^2) M^4 \nonumber \\ &&
+(-6 a^2 L^4 \pi u^10-240 a^2 L^4 \pi u^6-60 a^2 L^4 \pi u^8 \nonumber \\ &&
-192 a^2 \pi L^4-480 a^2 L^4 u^4 \pi-480 a^2 L^4 \pi u^2) M^2 \nonumber \\ &&
-8 a^3 L^6 u^6-24 a^3 L^6 u^4-32 a^3 L^6 u^2-16 a^3 L^6-a^3 L^6 u^8
\end{eqnarray}
\end{scriptsize}
Notice that the introduction of quantum gravity fluctuations makes
the thermodynamic stability of the radiating black hole more
complex. This is clear from the fact that the sign of the
determinant of the thermodynamic metric tensor depends on the
roots of $d^{c}(M,w=0)$. The notion of the thermodynamic
fluctuations is determined in terms of the intrinsic parameters,
\textit{viz.}, mass and radiation frequency of the constant
amplitude Hawking radiating black hole.

The limit $w=0$ shows that the sign of the determinant of the
metric tensor is a (negative) positive, signifying (un)stable
limiting thermodynamic fluctuations and thus a (un)favorable
condition for the stability of the Hawking radiating black hole.
Similarly, it is easy to notice that the sign of the scalar
curvature depends on the relative signs of $n^{c}(M,w=0)$ and
$d^{c}(M,w=0)$. More precisely, this indicates the nature of the
thermodynamic correlations on the parametric manifold
$(M_2(R),g)$. In the limit $w=0$, we find that the quantum
fluctuations make a complex behavior for an understanding of the
Planck size black holes. The present analysis shows that the
thermodynamic correlation length of such black holes is a
nontrivial function of the mass and radiation frequency. In the
limit $w=0$, notice further that the behavior of the correlation
length does not simplify sufficiently and it remains a non-trivial
function of the mass of the black hole.

As mentioned earlier, this section offers a quantum gravity
refined analysis of the fluctuation horizon analysis of the
previous section. As per the name of the consideration, we find
that the quantum gravity corrections indeed improve the
configuration to be well-behaved. However, the behavior of the
determinant of the metric tensor shows a similar qualitative
feature. We find that such limiting black holes become unstable
for small radiation frequency and large mass. For $u=1, m=1, L=1$,
the pictorial nature of the determinant of the metric tensor is
depicted in the Fig.[5]. We observe that the present system is
stable in the zero mass limit, but acquires a large negative
fluctuation of order $10^{6}$. As expected on physical ground, it
is worth mentioning that the limiting zero mass black hole becomes
highly self-contracting.

As a matter of fact, the corresponding Fig.[6] shows for $u=1,
m=1, L=1$ that there are four peaks of thermodynamic instability.
For small mass and small radiation frequency, we see in the above
limit of the parameters that the negative strength of the
thermodynamic interactions is of order $10^{5}$. Thus, the leading
order quantum gravity corrections improve the stability of the
radiation process. Fig.[6] shows further that the system acquires
a four triangular wall of instability, in the limit $\{(M,w)| \
\vert M \vert, \vert w \vert \rightarrow 0\}$. Physically, if one
shares the notion that the black hole has a remnant, then the
above picture changes slightly. We find that the small mass and
small frequency black holes have only four peaks of instability,
and secondly they are unstable with a relatively smaller degree of
self-interaction. Such a regulation of the instability is an
outcome of the quantum gravity corrections to the horizon, which
reduce the strength of the instability, when they are taken into
account.
\subsection{Decreasing Amplitude Fluctuations}

In this subsection, we explore the nature of an ensemble of
Hawking radiating decreasing amplitude black holes with the
logarithmic quantum corrections to the horizon area. To consider
the typical nature of the thermodynamic fluctuation, we choose the
radiating ensemble as the function of mass and radiation
frequency, taken as the system fluctuating parameters. The horizon
area is given by the following function of the mass and radiation
frequency
\begin{scriptsize}
\begin{eqnarray}
B(M,w) := 4 \pi (\frac{M}{m})^2-4 \pi \frac{a^2}{(1+16 M^2w^2)}
\end{eqnarray}
\end{scriptsize}
Thus, the quantum gravity corrected horizon area \cite{ZCZ} is given by
\begin{scriptsize}
\begin{eqnarray}
A(M,w) := (\frac{\pi}{L^2}) \bigg( (\frac{M}{m})^2- \frac{a^2}{(1+16 M^2 w^2)} \bigg)
+ a \ln \bigg( \frac{4 \pi}{L^2} (\frac{M}{m})^2- \frac{4 \pi}{L^2} \frac{a^2}{(1+16 M^2 w^2)} \bigg)
\end{eqnarray}
\end{scriptsize}
When both the parameters are allowed to fluctuate, we can exploit
the definition of the Hessian function $Hess(A(M,w))$ of the
Eqn.(\ref{metric}). Thus, we see for the decreasing amplitude
radiating black holes that there is the following set of
parametric pair correlations
\begin{scriptsize}
\begin{eqnarray} \label{QDmetric}
g_{MM} &=& \frac{2}{m^2 L^2}
\frac{( g^{dMM}_{14}M^{14} +g^{dMM}_{12} M^{12} +g^{dMM}_{10}M^{10} +g^{dMM}_8M^8 +g^{dMM}_6 M^6
+g^{dMM}_4 M^4 + g^{dMM}_2 M^2 +g^{dMM}_0)}{(M^2+16 M^4 w^2-a^2 m^2)^2 (1+16 M^2 w^2)^3}  \nonumber \\
 g_{Mw} &=& \frac{64 w M a^2}{L^2}
\frac{( g^{dMw}_{10} M^{10}+ g^{dMw}_{8} M^8+ g^{dMw}_{6} M^6+ g^{dMw}_{4} M^4+ g^{dMw}_{2} M^2
+ g^{dMw}_{0})}{(M^2+16 M^4 w^2-a^2 m^2)^2 (1+16 M^2 w^2)^3}  \nonumber \\
g_{ww} &=& \frac{32 M^2 a^2}{L^2}
\frac{( g^{dww}_{10}M^{10}+ g^{dww}_{8}M^{8}+ g^{dww}_{6}M^{6}+ g^{dww}_{4}M^{4}
+ g^{dww}_{2}M^{2}+ g^{MM}_{0})}{(M^2+16 M^4 w^2-a^2 m^2)^2 (1+16 M^2 w^2)^3}
\end{eqnarray}
\end{scriptsize}
where the functions $\{g^{ab}_{2i} \vert \ a,b \in \{ M,w \}; i
\in \{0,1,2,3,4,5,6,7\}\}$ are defined in the Appendix (B1).
Furthermore, we see that the thermodynamic fluctuations of the
quantum corrected decreasing amplitude radiating black holes
comply with the physical expectation that the system would
approach a stable configuration. In the limit of $g^{ab}_{2i} \ge
0$, we notice that the heat capacities, defined as the self-pair
correlations, remain positive quantities for a massive radiating
black hole. Moreover, an easy analysis shows the global properties
of the thermodynamic fluctuations. In the present case, we find
that the determinant of the metric tensor is given by
\begin{scriptsize}
\begin{eqnarray} \label{QDdet}
\Vert g(M,w) \Vert = -\frac{64 a^2 M^2}{m^2L^4}
\frac{\sum_{n=0}^{9} g^{d}_{2n} M^{2n}}{(M^2+16 M^4 w^2-a^2 m^2)^3(1+16 M^2 w^2)^5}
\end{eqnarray}
\end{scriptsize}
where the functions $\{g^{d}_{2n} \vert \ 0 \le n \le 9 \}\}$ are defined in the Appendix (B2).
\begin{figure}
\hspace*{0.5cm}
\includegraphics[width=8.0cm,angle=-90]{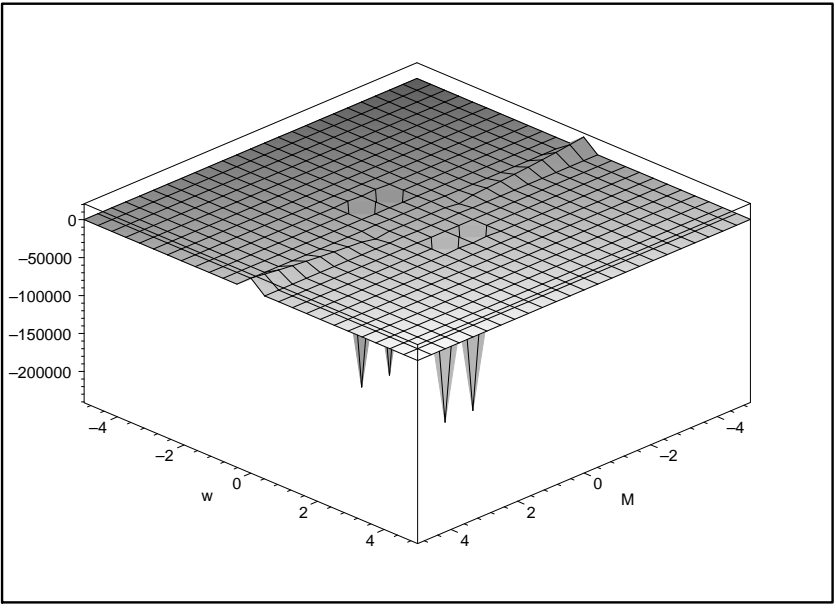}
\caption{The determinant of the metric tensor plotted as the function of the radiation
frequency $w$ and mass $M$, describing the fluctuations in the decreasing amplitude
Hawking radiating black hole with the quantum gravity corrections.}
\vspace*{0.5cm}
\end{figure}
As we have shown in the previous considerations, the determinant
of the metric tensor is non-zero for non-zero mass and
$\{g^{d}_{2n} \}$. It is worth mentioning that this configuration
is globally unstable over a large part of the intrinsic parametric
configurations. This is also intelligible from the fact that the
responsible quantum corrected entropy tends to its maximum value,
while the same culmination does not remain valid over the entire
domain of the Ruppenier's state-space manifold. For a class of
$g^{d}_{2n}$, we find that the configuration is stable in the
limit $M^2+16 M^4 w^2 < a^2 m^2$. For a given Planck mass, we
observe that the thermodynamic stability constraint requires a
specific choice of one or more of the system parameters,
\textit{viz.}, mass, frequency and amplitude of the radiation.

In order to examine important global properties in the Hawking
radiating black hole configurations, one is required to determine
the associated invariants of the underlying thermodynamic
geometry. As previously defined, the simplest invariant turns out
to be the intrinsic scalar curvature, which may be easily computed
by simply following our technology of the intrinsic Riemannian
geometry. Our analysis further discovers that there exists an
interacting thermodynamic configuration with the quantum gravity
contributions. For the present configuration, a little involved
computation shows that the Ricci scalar takes the following
expression
\begin{scriptsize}
\begin{eqnarray} \label{QDcur}
R(M,w) = (\frac{Lm}{M})^2
\frac{\sum_{n=0}^{18} r^{d}_{2n} M^{2n}}{(\sum_{n=0}^{9} g^{d}_{2n} M^{2n})^2}
\end{eqnarray}
\end{scriptsize}
where the functions $\{r^{d}_{2n} \vert \ 0\le n \le 18 \}$ are
defined in the Appendix (B3). Based on the Appendix (B2, B3), it
is worth emphasizing further that the thermodynamic scalar
curvature is non-vanishing, as long as the mass of the Hawking
radiating black hole is non-zero. The argument follows from the
fact that $r^{d}_{0}$ and $g^{d}_{0}$ are non-zero functions on
the parametric thermodynamic surface, for the entire domain of a
finite radiation frequency.
\begin{figure}
\hspace*{0.5cm}
\includegraphics[width=8.0cm,angle=-90]{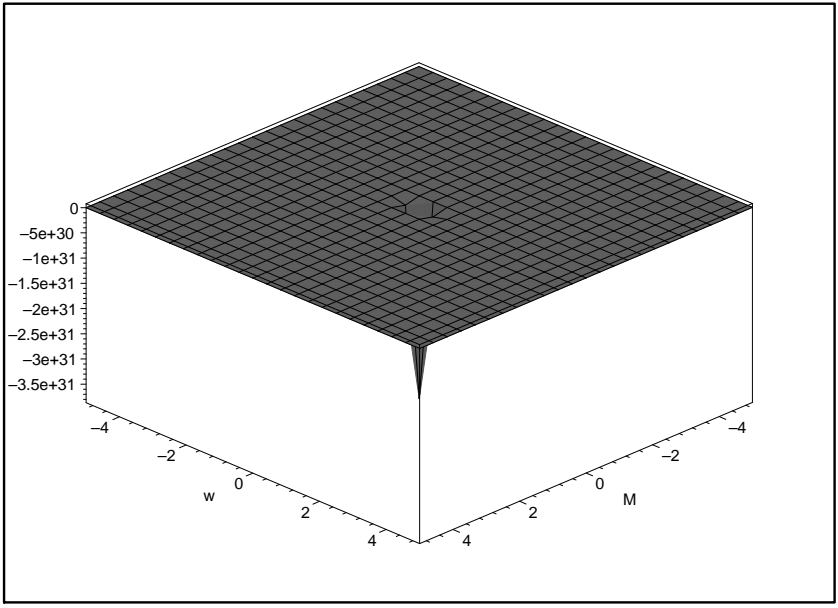}
\caption{The curvature scalar plotted as the function of the radiation frequency
$w$ and mass $M$, describing the fluctuations in the  decreasing amplitude
Hawking radiating black hole with the quantum gravity corrections.}
\vspace*{0.5cm}
\end{figure}
In the limit $w=0$, the local pair correlation functions reduce to
a set of simple values. Subsequently, we find that the components
of the covariant metric tensor are given by
\begin{scriptsize}
\begin{eqnarray}
g_{MM}&=& \frac{2}{m^2L^2}
\frac{(-\pi M^4+(a m^2 L^2+2 \pi a^2 m^2) M^2+a^3 m^4 L^2-\pi a^4 m^4)}{(M^2-a^2 m^2)^2}   \nonumber \\
g_{Mw}&=& 0   \nonumber \\
g_{ww}&=& \frac{32 M^2 a^2}{L^2} \frac{(-\pi M^4+(-a m^2 L^2+2 \pi a^2 m^2) M^2+a^3 m^4 L^2-\pi a^4 m^4)}{(M^2-a^2 m^2)^2}
\end{eqnarray}
\end{scriptsize}
It is easy to observe that the respective system becomes unstable
in the vanishing frequency limit. This follows from the negativity
of the limiting determinant of the metric tensor
\begin{scriptsize}
\begin{eqnarray}
\Vert g(M,w=0) \Vert = -\frac{64 a^2M^2}{m^2 L^4}\frac{d^{d}(M,w=0)}{(M^2-a^2 m^2)^3}
\end{eqnarray}
\end{scriptsize}
where the numerator $d^{d}(M,w=0)$ is defined by the following polynomial
\begin{scriptsize}
\begin{eqnarray}
d^{d}(M,w=0)&:=&-\pi^2 M^6+\pi^2 a^6 m^6+a^4 m^6 L^4-3 \pi^2 a^4 m^4 M^2+3 \pi^2 M^4 a^2 m^2 \nonumber \\ &&
+a^2 m^4 L^4 M^2+2 a^3 m^4 L^2 \pi M^2-2 a^5 m^6 L^2 \pi
\end{eqnarray}
\end{scriptsize}
It is worth mentioning that the thermodynamic behavior of the
decreasing amplitude Hawking radiating black hole is very
intriguing, even in the limit of the zero radiation frequency.
This follows from the fact that the stability of the limiting
small frequency configuration over the Gaussian fluctuations is
determined by the following quadratic and cubic inequalities
\begin{scriptsize}
\begin{eqnarray}
d^{d}(M^2,w=0)&>0& \ for \ M^2-a^2 m^2< 0, \nonumber \\
d^{d}(M^2,w=0)&<0& \ for \ M^2-a^2 m^2> 0
\end{eqnarray}
\end{scriptsize}
Thus, it is immediate to find the regions in which the small
frequency decaying amplitude Hawking radiating black hole is
(un)stable. Furthermore, the global nature of phase transitions,
if any, could be observed from the singularities in the limiting
scalar curvature. Specifically, in the limit $w=0$, we find that
the associated scalar curvature reduces to the following limiting
value
\begin{scriptsize}
\begin{eqnarray}
R(M,w=0) = \frac{L^2m^2}{M^2} \frac{n^{d}(M,w=0)}{d^{d}(M,w=0)^2}
\end{eqnarray}
\end{scriptsize}
where the polynomial in the numerator is given by
\begin{scriptsize}
\begin{eqnarray}
n^{d}(M,w=0)&:=&-\pi^3 M^{12}+(2 a \pi^2 m^2 L^2+6 a^2 \pi^3 m^2) M^{10} \nonumber \\ &&
+(-15 a^4 \pi^3 m^4+a^2 \pi m^4 L^4-a^3 \pi^2 m^4 L^2) M^8 \nonumber \\ &&
+(-12 a^5 \pi^2 m^6 L^2-2 a^4 \pi m^6 L^4+20 a^6 \pi^3 m^6) M^6 \nonumber \\ &&
+(-a^5 m^8 L^6-6 a^6 \pi m^8 L^4-15 a^8 \pi^3 m^8+22 a^7 \pi^2 m^8 L^2) M^4 \nonumber \\ &&
+(6 a^10 \pi^3 m^10+10 a^8 \pi m^10 L^4-14 a^9 \pi^2 m^10 L^2-2 a^7 m^10 L^6) M^2 \nonumber \\ &&
+3 a^11 \pi^2 m^12 L^2-3 a^10 \pi m^12 L^4-a^12 \pi^3 m^12+a^9 m^12 L^6
\end{eqnarray}
\end{scriptsize}
Herewith, we find that the decreasing amplitude radiation black
holes with an incorporation of the quantum correction possess
relatively distinct thermodynamic behavior over the parametric
fluctuations. The small frequency decaying amplitude Hawking
radiating black hole is non-interacting at the real roots of a six
degree equation $n^{d}(M^2,w=0)=0$. As in the case of the constant
amplitude Hawking radiating black hole, we observe that the
introduction of quantum gravity fluctuations gives the decaying
amplitude Hawking radiating black hole a more complex
thermodynamic stability character. In fact, for such decaying
amplitude Hawking radiating black holes, this is clear from the
fact that the sign of the determinant of the thermodynamic metric
tensor depends on the roots of $d^{d}(M^2,w=0)$, even far way from
the Planck size black holes. However, apart from the computational
complexities, the conceptual consideration is very clear and lucid
from the intrinsic geometric perspective. Thus, the notion of the
thermodynamic fluctuations is determined in terms of the intrinsic
parameters, \textit{viz.}, mass and radiation frequency of the
decaying amplitude Hawking radiating black hole with an account of
the quantum gravity.

In the small frequency limit, \textit{i.e.} $w=0$, it is evident
that the determinant of the metric tensor is negative (positive),
signifying (un)stable limiting thermodynamic fluctuations and thus
a (un)favorable condition for the stability of the decaying
amplitude Hawking radiating black hole. Notice further that the
sign of the scalar curvature depends on the relative signs of
$n^{d}(M,w=0)$ and $d^{d}(M,w=0)$. In the limit $w=0$, their
identical signs indicate that the nature of the thermodynamic
correlations is repulsive on the parametric manifold $(M_2(R),g)$.
We find further that the quantum fluctuations produce a complex
behavior of the decaying amplitude Hawking radiating black holes.
Even in the limit $w=0$, the correlation length of the decaying
amplitude Hawking radiating black holes is non-trivially dependent
on the mass of the black hole.

For $a=1, m=1, L=1$, the graphical behavior of the determinant of
the metric tensor is shown in the Fig.[7]. We observe that such
limiting black holes become unstable in the limit of small
frequency and small mass. For $a=1, m=1, L=1$, the pictorial
nature of the scalar curvature is depicted in the Fig.[8].
Herewith, we notice that the system is unstable in the zero mass
limit, but acquires a large negative thermodynamic fluctuation of
order $10^{5}$ for the determinant of the metric tensor, and
$10^{31}$, for the thermodynamic scalar curvature. From the
Fig.[7,8], we observe under quantum gravity contributions that the
global nature of the decaying amplitude Hawking radiating black
hole is very similar to the constant amplitude Hawking radiating
black hole. As per the present consideration, the Fig.[8] shows
that the limiting zero mass and zero frequency black hole becomes
highly self-contracting as in the case of a constant amplitude
black hole. It is worth mentioning that the limit of vanishing
mass and radiation frequency makes the negative strength of the
thermodynamic global interaction to be of higher order, in
comparison with both the constant amplitude black hole, with and
without the quantum gravity corrections, and only decreasing
amplitude horizon fluctuating black hole.

In the process of the quantum gravity corrections to the horizon,
we observe that the Fig.[8] has the same qualitative behavior as
the corresponding Fig.[4] with $a=1, m=1$. This shows that the
thermodynamic system is almost insensitive to the quantum gravity
contributions, as they alter the global thermodynamic interaction
from an order of $10^{30}$ to $10^{31}$. Whilst, the behavior of
the determinant of the metric tensor is quite different, and thus
it depends significantly of the quantum parameter $L$. In fact, we
find that the positive metric fluctuations are diminished by the
quantum gravity fluctuations. However, they introduce four
negative peaks of the thermodynamic conical singularities of the
strength $10^{5}$. Thus, in the limit $\vert M \vert \rightarrow
0$, the determinant of the metric tensor acquires four bumps of
small strength thermodynamic instabilities, while the scalar
curvature acquires a large conical thermodynamic instability of
the strength $10^{31}$. As a matter of fact, the existence of the
black hole remnant would again make the Fig.[7,8] to be
capped-off, and thus the local thermodynamic singularity would be
regulated to be four large finite asymmetric cylinders, while the
underlying global thermodynamic singularity would be regulated to
a large finite asymmetric cylinder.

Our analysis illustrates that the physical properties of the
specific local and global thermodynamic correlations may easily be
exactly exploited in general, without any approximation. Within a
small neighborhood of proclaimed statistical fluctuations over an
equilibrium ensemble of configurations, the framework of the
thermodynamic geometry explicates the functional nature of
parameters without any surprise. As mentioned in the Appendix-I,
the explicit forms of the thermodynamic correlations are
well-behaved functions of the mass and the radiation frequency of
the considered Hawking radiating black hole. The definite behavior
of correlations, as accounted in the previous section, and the
leading order quantum gravity logarithmic corrections in present
section, suggest that the possible constant and decaying amplitude
Hawking radiating black holes do have definite stability
properties, except that the determinant may be non-positive
definite in some specific limit of the mass and radiation
frequency. As mentioned before, it is worth recalling that these
black hole are generically well-defined from the thermodynamic
perspective and indicate an interacting statistical basis in
general. Herewith, we discover that the thermodynamic geometry
indicate a promising nature of the Hawking radiating black hole
configurations, with and without the leading order quantum gravity
contributions.
\section{Hawking Radiation Energy Flux}
In the present section, we explore the nature of an ensemble of
generic Hawking radiating black holes generated with the variable
mass and variable radiation frequency of the black hole. To
consider the most general case, we choose the energy flux as the
function of mass and frequency. When both of them are allowed to
fluctuate, we can exploit the definition of the Hessian function
$Hess(F(M,w))$ of the energy flux of the Hawking radiation. As
explored in the previous sections, we specialize ourselves to the
constant amplitude and decaying amplitude radiating black holes.
\subsection{Constant Amplitude Fluctuations}
In this section, we study the case of a constant amplitude (u) Hawking radiation.
Subsequently, the energy flux of the constant amplitude Hawking radiation is given by
\begin{scriptsize}
\begin{eqnarray}
F(M,w)=\frac{1}{768 \pi M^2} (1+\frac{4 u^2 \pi M w}{\exp(8 \pi M w)-1})
\end{eqnarray}
\end{scriptsize}
As per the consideration of the present investigation, we treat
the $x^a = (M, w) \in M_2$ to be fluctuating thermodynamic
variables. The consideration of the intrinsic geometry is realized
by defining the metric tensor as the Hessian matrix of the energy
flux of the Hawking radiating configuration. In the present
subsection, we treat the quantum gravity amplitude ($u$) to be a
constant. From the viewpoint of the thermodynamic limit, our
analysis shares the fact that the equilibrium of the Hawking
radiated particles and remnant(s) black hole, if any, should be
extracted from the limiting equilibrium statistical
configuration(s). Of course, the procedure of the extraction could
in general be highly non-trivial. However, the present
consideration leads us to the conclusion that the energy flux of
the radiation defines a set of radially computable expression
against both the local and global thermodynamic correlations.
First of all, such a conclusion arises from the Hessian matrix of
the energy flux fluctuations. Explicitly, the components of
covariant metric tensor are given by
\begin{scriptsize}
\begin{eqnarray} \label{RCmetric}
g_{MM} &=& \frac{1}{384\pi M^4} \frac{(c^{cMM}_3 M^3+c^{cMM}_2 M^2 +c^{cMM}_1 M+ c^{cMM}_0)}{(\exp(8 \pi M w)-1)^3} \nonumber \\
g_{Mw} &=&  \frac{ u^2}{192 M^2} \frac{( c^{cMw}_2 M^2 +c^{cMw}_1 M -c^{cMw}_0)}{(\exp(8 \pi M w)-1)^3} \nonumber \\
g_{ww} &=& \frac{\pi u^2}{12} \exp(8 \pi M w) \frac{(c^{cww}_1 M-c^{cww}_0)}{(\exp(8 \pi M w)-1)^3}
\end{eqnarray}
\end{scriptsize}
For the constant amplitude Hawking radiating black holes, the
above behavior of the parametric pair correlations shows that the
heat capacities, defined as the self-pair correlations, are
positive in a domain of the functions $\{c^{cab}_i \vert \ a,b \in
\{ M,w \}; i \in \{0,1,2,3\}\}$. The precise functional nature of
$\{c^{cab}_i \vert \ a,b \in \{ M,w \}; i \in \{0,1,2,3\}\}$ is
given in the Appendix (C1). Subsequently, we find that the
determinant of the metric tensor is
\begin{scriptsize}
\begin{eqnarray} \label{RCdet}
\Vert g(M,w) \Vert = \frac{u^2}{36864M^4} \frac{(h^{c}_3M^3+h^{c}_2M^2+h^{c}_1M+h^{c}_0)}{(\exp(8 \pi M w)-1)^5}
\end{eqnarray}
\end{scriptsize}
where the functions $\{h^{c}_0,h^{c}_1,h^{c}_2,h^{c}_3\}$ are presented in the Appendix (C2).
\begin{figure}
\hspace*{0.5cm}
\includegraphics[width=8.0cm,angle=-90]{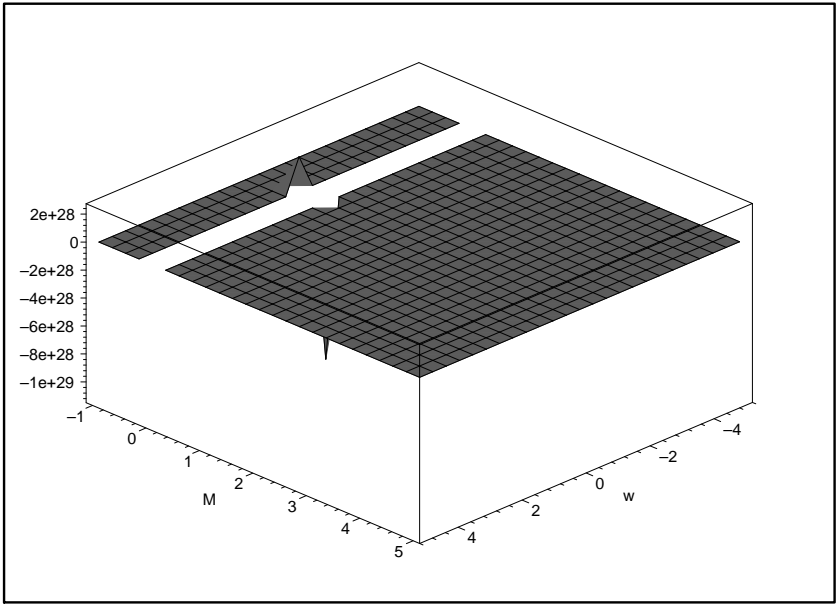}
\caption{The determinant of the metric tensor plotted as the function of the radiation
frequency $w$ and mass $M$, describing the energy flux fluctuations in the constant amplitude
Hawking radiating black hole.}
\vspace*{0.5cm}
\end{figure}
It is not difficult to compute an exact expression for the scalar
curvature describing the global parametric intrinsic correlations.
By defining a set of radiation functions, we find explicitly that
the most general scalar curvature takes the following rational
form
\begin{scriptsize}
\begin{eqnarray} \label{RCcur}
R(M,w) = 6144  \pi M^2 \exp(8 \pi M w) \frac{(l^{c}_5M^5+ l^{c}_4M^4+ l^{c}_3 M^3
+ l^{c}_2 M^2+ l^{c}_1 M+ l^{c}_0)}{(h^{c}_3M^3+h^{c}_2M^2+h^{c}_1M+h^{c}_0)^2}
\end{eqnarray}
\end{scriptsize}
where the functions $\{l^{c}_0,l^{c}_1,l^{c}_2,l^{c}_3,l^{c}_4,l^{c}_5\}$ are defined in the Appendix (C3).
\begin{figure}
\hspace*{0.5cm}
\includegraphics[width=8.0cm,angle=-90]{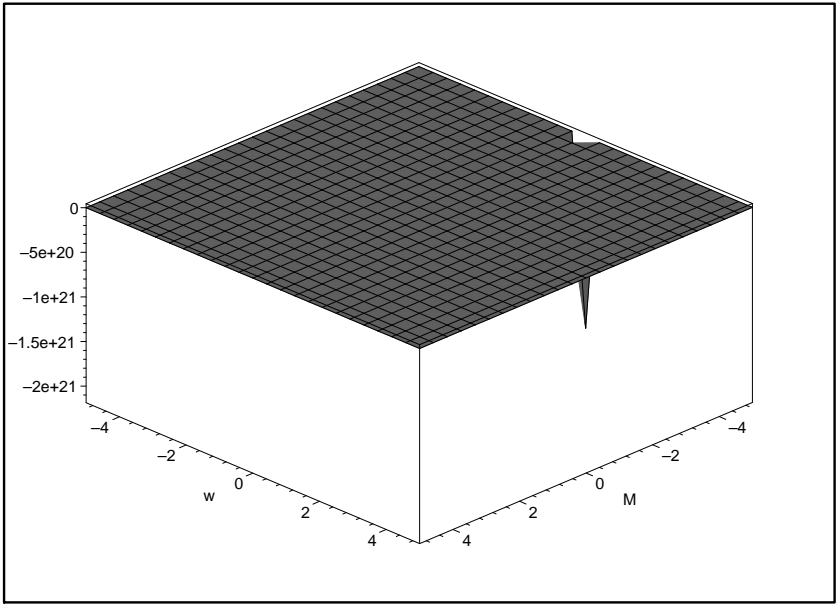}
\caption{The curvature scalar plotted as the function of the radiation frequency
$w$ and mass $M$, describing the energy flux fluctuations in the constant amplitude
Hawking radiating black hole.}
\vspace*{0.5cm}
\end{figure}
Notice that all the functions $\{h^{c}_i\}$ vanish identically for
$w=0$. From the viewpoint of thermodynamic stability, the
vanishing of the combination
$h^{c}_3M^3+h^{c}_2M^2+h^{c}_1M+h^{c}_0$ makes the static
frequency configuration ill-defined. The determinant of the
thermodynamic metric tensor vanishes and the Ricci scalar diverges
in the limit of zero frequency. This shows that the configuration
may suffer certain phase transitions. Thus, the analysis of the
constant amplitude configuration anticipates that the fluctuations
of the energy flux would not stop, once the black hole starts
Hawking radiating.

Pictorially, the behavior of the determinant of the metric tensor
is shown in the Fig.[9]. We anticipate that the graph has a
significant change in the regime of small mass and small radiation
frequency. In the limit of small radiation frequency and large
mass, we find that such a limiting black hole becomes highly
unstable and shows a large negative strength thermodynamic
instability of order $10^{21}$. For a unit amplitude of the
constant Hawking radiation $u=1$, the pictorial nature of the
scalar curvature is depicted in the Fig.[10]. Herewith, we notice
that the system is unstable in the zero mass and zero frequency
limit. Interestingly, the graphical analysis of the Fig.[9] shows
that the constant amplitude Hawking radiating black hole acquires
a region of small positive thermodynamic fluctuations in the limit
$\{ M \rightarrow 0^{+}, w \rightarrow 0^{+} $, while it acquires
a region of negative fluctuation in the limit $\{ M \rightarrow
0^{-}, w \rightarrow 0^{-} $. As expected from the physical basis
of the Hawking radiating black hole, it is worth mentioning that
the limiting zero mass and zero frequency black hole becomes
highly self-contracting. As a matter of fact, the Fig.[10] shows
that the limit of vanishing frequency yields a negative strength
of the thermodynamic interaction of order $10^{21}$. In this
process. we observe further that the system acquires a spike of
thermodynamic instability, in the limit $\vert w \vert \rightarrow
0$.

Physically, if one takes an account of the fact that the black
holes have a remnant, then the above picture changes slightly and
we find herewith that the Hawking radiating black holes are
unstable with a high degree of finite self-interaction. A refined
analysis would be the matter of future investigation, where one
might take an account of the higher loop quantum gravity
corrections arising from the path integral formulation of the
gravity, or the ones which are originating from the gravitational
aspect of the string theory, with higher order correction in the
string length and Planck length. Thus, an appropriate notion of
string theory or quantum gravity theory may offer a better
thermodynamic understanding of the Hawking radiating black holes.
Specifically, it is important to mention that the underlying
statistical correlation length of the underlying ensemble reducing
to a nearly equilibrium system requires an understanding of the
full theory of the quantum gravity. The present investigation is
however premature, for showing such atypical qualification of the
Hawking radiating black hole.
\subsection{Decreasing Amplitude Fluctuations}
In this section, the case of the decaying amplitude York model
\cite{York} is analyzed. In order to take a closer look at the
thermodynamic limit of the energy flux of a Hawking radiating
black hole, we consider the thermodynamic geometry of the limiting
equilibrium ensemble of the black holes. To be precise, we
consider the implications of the York model correction to the
thermodynamic geometry. With such a consideration, the energy flux
of the Hawking radiation with a decaying amplitude
\begin{scriptsize}
\begin{eqnarray}
u:= a \frac{m}{M}
\end{eqnarray}
\end{scriptsize}
is given by
\begin{scriptsize}
\begin{eqnarray}
F(M,w) := \frac{1}{768 \pi M^2} (1+ \frac{4 \pi a^2 m^2 w}{M (\exp(8 \pi M w)-1)})
\end{eqnarray}
\end{scriptsize}
The metric tensor of the thermodynamic geometry in the energy flux
representation may be obtained as before from the Hessian matrix
of the energy flux of the Hawking radiation. With respect to
previously acclaimed thermodynamic variables, the thermodynamical
behavior of the decaying amplitude radiation flux is not difficult
to analyze. Furthermore, it is legitimate to expect that the local
correlations would acquire a similar structure. In this
subsection, we precisely show that the thermodynamic correlations
take a set of the similar expressions as for the constant
amplitude energy flux. Explicitly, as the function of the
radiation frequency and mass of the black hole, we find that the
components of the covariant metric tensor are given by
\begin{scriptsize}
\begin{eqnarray} \label{RDmetric}
g_{MM} &=& \frac{1}{384 \pi M^5} \frac{(c^{dMM}_2M^2+ c^{dMM}_1M+ c^{dMM}_0)}{(\exp(8 \pi M w)-1)^3} \nonumber \\
g_{Mw} &=& \frac{a^2m^2}{192M^4} \frac{(c^{dMw}_2M^2 +c^{dMw}_1 M- c^{dMw}_0)}{(\exp(8 \pi M w)-1)^3}  \nonumber \\
g_{ww} &=& \frac{\pi a^2 m^2}{12 M^2} \exp(8 \pi M w) \frac{(c^{dMw}_1M+ c^{dMw}_0)}{(\exp(8 \pi M w)-1)^3}
\end{eqnarray}
\end{scriptsize}
where the functions $\{c^{dab}_i \vert \ a,b \in \{ M,w \}; i \in \{0,1,2\}\}$ given in the Appendix (D1).
Notice further that the principle components of the metric tensor signifying self pair correlations,
are positive definite functions over a range of the parameters. Physically, this signifies a set of
heat capacities against the intrinsic interactions arising from the mass and frequency of the radiation.
Over the Gaussian limit such correlations form stable correlations, if the determinant of the metric tensor
takes a positive value on the parametric surface of the Hawking radiation. For the given energy flux,
it is evident that the determinant of the metric tensor may easily be presented in the following form
\begin{scriptsize}
\begin{eqnarray} \label{RDdet}
\Vert g(M,w) \Vert = \frac{a^2 m^2}{36864 M^8 } \frac{(h^{d}_3 M^3 +h^{d}_2 M^2 +h^{d}_1 M+h^{d}_0)}{(\exp(8 \pi M w)-1)^5}
\end{eqnarray}
\end{scriptsize}
where the functions $\{h^{d}_0,h^{d}_1,h^{d}_2,h^{d}_3 \}$ are
defined in the Appendix (D2).
\begin{figure}
\hspace*{0.5cm}
\includegraphics[width=8.0cm,angle=-90]{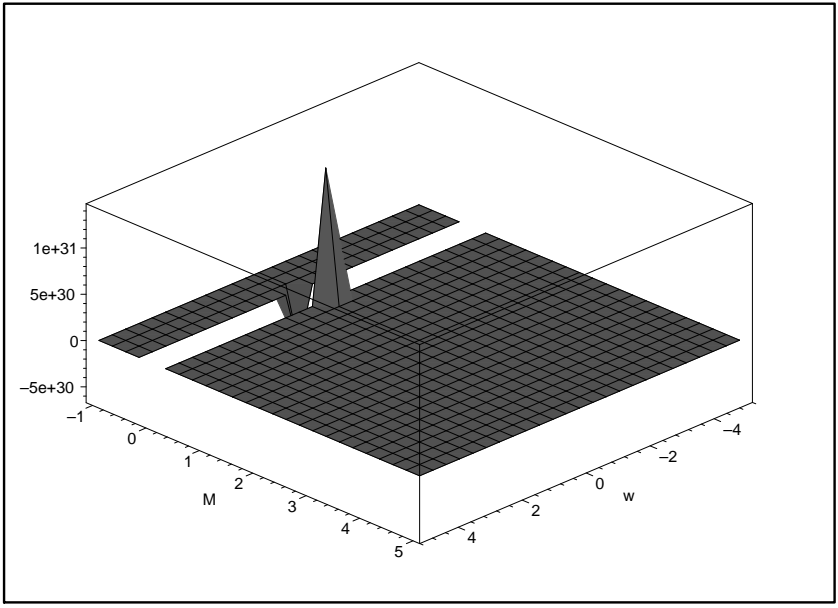}
\caption{The determinant of the metric tensor plotted as the function of the radiation
frequency $w$ and mass $M$, describing the energy flux fluctuations in the decreasing amplitude
Hawking radiating black hole.}
\vspace*{0.5cm}
\end{figure}
As a standard interpretation, the thermodynamic scalar curvature
describes the nature of underlying statistical interactions of a
possible particle on the horizon configurations of the black hole.
In particular, it turns out to be non-zero and well defined for a
wide range of the parameters of the Hawking radiating black hole
solution. Furthermore, the intrinsic scalar curvature
corresponding to the underlying thermodynamic geometry elucidates
the typical feature of Gaussian fluctuations about the limiting
equilibrium particles of the desired mass and radiation frequency.
This picture is evident from the consideration of the quantum
field theory on curved space-time, implying a modified asymptotic
spectrum. More precisely, one can consider ingoing and outgoing
waves and compute the properties of the Hawking radiated particles
from the Bogoliubov coefficients and thus the spectrum of the
Hawking radiation energy flux. As in the case of constant
amplitude radiation, we find further for the decaying amplitude
radiation that the Ricci scalar curvature reduces to the following
specific expression
\begin{scriptsize}
\begin{eqnarray} \label{RDcur}
R(M,w) = 12288 \pi M^4 \exp(8 \pi M w) \frac{(l^{d}_5 M^5 +l^{d}_4 M^4 +l^{d}_3 M^3 +l^{d}_2 M^2 +l^{d}_1 M
+l^{d}_0)}{(h^{d}_3 M^3 +h^{d}_2 M^2 +h^{d}_1 M+ h^{d}_0)^2}
\end{eqnarray}
\end{scriptsize}
where the functions $\{l^{d}_0,l^{d}_1,l^{d}_2,l^{d}_3,l^{d}_4,l^{d}_5\}$ are relegated to the Appendix (D3).
\begin{figure}
\hspace*{0.5cm}
\includegraphics[width=8.0cm,angle=-90]{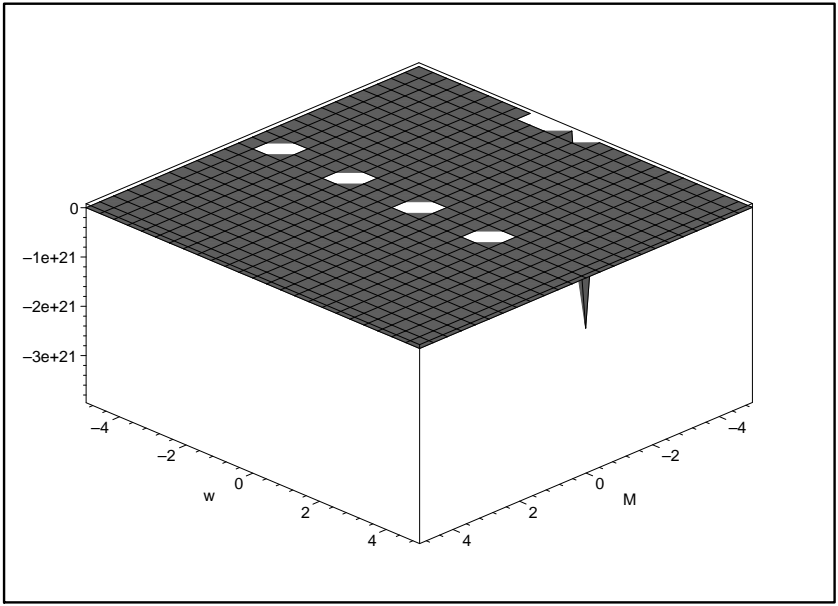}
\caption{The curvature scalar plotted as the function of the radiation frequency
$w$ and mass $M$, describing the energy flux fluctuations in the decreasing amplitude
Hawking radiating black hole.}
\vspace*{0.5cm}
\end{figure}
In this case, it turns out that the observation made in the
constant amplitude case continues to hold. Specifically, we find
that all the functions $\{h_i\}$ vanish identically for $w=0$. As
mentioned before in the previous section, the vanishing of the
combination $h^{d}_3M^3+h^{d}_2M^2+h^{d}_1M+h^{d}_0$ makes the
static frequency configuration ill-defined, even for a decreasing
amplitude Hawking radiating black hole.

From the viewpoint of thermodynamic stability, the determinant of
the thermodynamic metric tensor vanishes and the Ricci scalar
diverges in the limit of zero frequency, which implies that the
system becomes unstable. In fact, such an analysis suggests that
the configuration may suffer a phase transition, as well. Thus, as
in the case of the constant amplitude configuration, the
decreasing amplitude analysis anticipates that the fluctuations of
the energy flux should not stop, once the black hole starts
Hawking radiating. This opens an avenue to understand the nature
of the interaction, while a black hole is approaching the limiting
thermodynamic configuration.

As in the case of the constant amplitude Hawking radiation energy
flux arising from the two parameter black hole, we observe in the
case of a decreasing amplitude Hawking radiating energy flux that
the behavior of the determinant of the metric tensor remains
almost the same for $a=1, m=1$. Pictorially, this is shown in the
Fig.[11], along with the fact that above observation follows from
a direct comparison of the Fig.[11] with that of the Fig.[9]. For
unit amplitude of decreasing Hawking radiating black hole $u=1$
and decaying amplitude black hole, an observation follows from the
anticipation of the qualitative properties of the Figs.[10] and
Fig.[12]. We notice that both graphs of the scalar curvatures have
a significant change along the line of $M=0$. The regime of small
radiation frequency remains almost the same, except for the fact
that the small radiation frequency and large mass region becomes
relatively narrower. In this regime, we find that the limiting
black hole becomes highly unstable and it possesses a large
negative strength thermodynamic instability of order $10^{21}$, as
per the case of the constant amplitude amplitude energy flux
Hawking radiating black hole.

Globally speaking, the pictorial nature of the scalar curvature,
as depicted in the Fig.[12], captures non-perturbative quantum
gravity contributions, and thus the parametric surface has a
different topology. Herewith, we notice that the system is
unstable in the large mass and zero frequency limit. As in the
case of the constant amplitude Hawking radiating black hole, the
graphical analysis of the Fig.[11] shows that the decreasing
amplitude Hawking radiating black hole acquires a region small
positive thermodynamic fluctuations in the limit of $\{ M
\rightarrow 0^{+}, w \rightarrow 0^{+} \}$, while it acquires a
region of negative fluctuation in the limit of $\{ M \rightarrow
0^{-}, w \rightarrow 0^{-} \}$. As expected from physical basis of
the Hawking radiating black hole, the limiting zero mass and zero
frequency black hole becomes highly self-contracting and much more
intricate, for decreasing amplitude Hawking radiating black hole.
As a matter of fact, the Fig.[12] shows that the limit of
vanishing frequency makes a negative strength of the thermodynamic
interaction of order $10^{21}$. In this process. we observe
further that the system acquires a double cut spike of
thermodynamic instability, in the limit $\vert w \vert \rightarrow
0$.

As depicted in the Appendix-II, it is worth stating that the
mathematical behavior of the thermodynamic metric tensor is given
by the Hessian matrix of the concerned energy flux. In the case of
the fluctuation of the energy flux, the mass and the radiation
frequency of the Hawking radiating black hole form the co-ordinate
chart on the intrinsic manifold of the parameters, which
characterize the intrinsic geometric thermodynamic correlations.
Our explicit computation shows that the exact set of correlations
for both the constant amplitude and decaying amplitude energy flux
of a Hawking radiating black holes is realized by employing the
intrinsic geometric notion. Herewith, we have shown that the
limiting thermodynamic correlations arising from the fluctuations
of the energy flux demonstrate a simple structure in both cases,
\textit{viz,}, constant and decreasing amplitude Hawking radiating
configurations. In a small neighborhood introduced over an
ensemble of equilibrium configurations, the functional nature of
parametric fluctuations may be exactly computed as the function of
the radiation frequency and mass, with the form $\exp(8 \pi Mw)$.

Importantly, if one takes an account of the quantum gravity
effects that the black holes would have remnant, and thus the
above picture would change slightly, and one would have a
rectified comprehension of the Hawking radiating black holes. This
may offer a microscopic origin why such configurations are
unstable with a high degree of finite self-attraction. A refined
analysis would be the matter of future investigation, where one
might take an account of the higher loop quantum gravity
corrections arising from the path integral formulation of the
gravity, beyond the saddle point, or the ones originating from the
gravitational aspect of the string theory or D-brane ensemble,
with higher order string length and Planck length contributions to
the radiation energy flux. Thus, an appropriate notion of string
theory or quantum gravity theory may offer a better thermodynamic
understanding of the York type \cite{York} decreasing amplitude
Hawking radiating black holes. Specifically, it is important to
mention that the underlying statistical correlation length of the
underlying ensemble of decreasing amplitude Hawking radiating
black holes might be anticipated to reduce to a nearly equilibrium
system. Such unified issues are beyond the scope of the present
investigation to fully anticipate the microscopic comprehension of
the Hawking radiation, quantum gravity and black hole physics.
\section{Conclusion and Outlook}

Intrinsic geometric technology is well-suitable for analyzing the
stability structure of Hawking radiation black holes, under the
fluctuations of the radiation parameters. Such fluctuations are
expected to arise due to non-zero heating effects, chemical
reactions and possible conventional instabilities associated with
the Hawking radiation, fluctuating horizon and quantum gravity
corrections to the black hole configuration. The intrinsic
geometric method is thus used to investigate the structure of the
constant amplitude and decaying York amplitude ensembles, designed
from the consideration of background space-time fluctuations. The
present analysis is well suited for the statistical understanding
of the formation of an equilibrium physical black hole. The
quantum gravity characterization procedure is presented for both
the constant amplitude radiation and decaying amplitude Hawking
radiating black holes. The respective considerations are
pictorially depicted for a black hole of (i) unit amplitude
constant Hawking radiation and (ii) unit York strength decaying
amplitude Hawking radiation.

In both types of horizon fluctuations, it turns out that the
Hawking radiating black hole corresponds to an interacting
statistical system. The global stability of the ensemble is
determined from the positivity of the determinant of the metric
tensor. In both of the above cases, this reduces to the positivity
of an associated cubic equations for the entire domain of the mass
and the frequency of Hawking radiation. In both cases, we find
that the underlying configurations are globally non-interacting on
the real roots of degree 10 and degree 12 equations in the mass,
respectively. In the zero frequency limit, we observe further that
both the constant amplitude and decaying amplitude Hawking
radiating black holes are attractive from the thermodynamic
perspective. However, the first one having a constant amplitude
becomes unstable in the static frequency domain, while the second
one does not. The above observation follows from the fact that the
surface of the parameters has a negative determinant of the metric
tensor for the first consideration, while it takes a positive
definite value for the second case. Our observation thus supports
that the decreasing amplitude York model \cite{York} is a
preferable choice, yielding an attractive and thermodynamically
stable black hole.

Such an introduction is brought out from the consideration of
background space-time fluctuations, yielding fluctuating mass and
radiation frequency as the parameters for our intrinsic geometric
analysis. In fact, the fluctuating horizon geometry shows an
interesting property that the thermodynamic correlations are only
the even function of the mass of the black hole. The conclusion is
invariant, whether one considers the case of a constant amplitude
radiation or the decreasing amplitude York type model. Our
observation remains the same even after the incorporation of the
non-perturbative logarithmic quantum gravity corrections to the
chosen black hole horizon configurations. This shows the
robustness of the present analysis. In the case of the fluctuating
horizon configuration, the above robustness property follows from
the Eqns.(\ref{CCmetric},\ref{CCdet},
\ref{CCcur},\ref{CDmetric},\ref{CDdet},\ref{CDcur}).
Interestingly, it is surprising to note that the corresponding
observations remain intact under the quantum gravity logarithmic
corrections in all possible domains of the mass and radiation
frequency of the Hawking radiating black holes. It is  evident
from the corresponding quantum corrected thermodynamic metric
tensors, determinants of the associated metric tensor and the
respective scalar curvatures, \textit{viz,},
Eqns.(\ref{QCmetric},\ref{QCdet},\ref{QCcur},\ref{QDmetric},
\ref{QDdet},\ref{QDcur}).

In the dual energy flux representation, the nature of the generic
thermodynamic correlations over an ensemble of Hawking radiating
black hole, generated from the consideration of variable mass and
variable radiation frequency, yields an expected statistical
picture. Specifically, the present matter of intrinsic geometric
affair demonstrates that the limiting thermodynamic correlations
of a Hawking radiating black hole is well-defined in the frequency
domain $w \in (0, \infty)$. To investigate the most general
behavior of the fluctuations of the radiation energy flux, when
both the mass and the frequency are allowed to fluctuate, the
definition of Hessian function $Hess(F(M,w))$ of the energy flux
$F(M,w)$ offers the limiting thermodynamic characterization of a
Hawking radiating black hole. The exact local and global
thermodynamic behavior of the parametric correlations are evident
from the Eqns.(\ref{RCmetric},\ref{RCdet}, \ref{RCcur},
\ref{RDmetric}, \ref{RDdet},\ref{RDcur}). Following these
observations, the intrinsic geometric investigation concludes that
the Hawking radiating black holes are thermodynamically stable in
a specific domain of the radiating parameters and are interacting
for both the constant and York type decaying amplitude. This
offers an interesting platform to explore unified picture of the
horizon fluctuations, background space-time quantum fluctuations
and quantum statistical or limiting thermodynamic fluctuations.

With this understanding, our model is well suited and robust for
the Hawking radiating black holes with and without quantum gravity
effects. Such configurations are very popular nowadays because of
their best compliance with the vacuum difficulties and
renormalization group complications. These technicalities are
prone to arise from the structure of the infalling and outgoing
solutions of the D'Alembertian equations associated with the
background fluctuations, \textit{viz.} Vaidya space-time
geometries. From the above viewpoints, our intrinsic geometric
approach is thus very lucrative. It is worth mentioning that the
intrinsic parametrization principle is rapidly growing in robust
statistical theory and black hole physics, in the recent years. In
fact, the present analysis is capable of accommodating the
corresponding quantum considerations, as well. The present paper
has explored these notions in detail and has additionally offered
the associated pictorial depictions of the intrinsic geometric
invariants for the unit Planck length, unit constant amplitude,
unit York strength and unit logarithmic quantum gravity correction
coefficient. In such considerations, the exact mathematical
limiting formulae are obtained for the corresponding thermodynamic
characterization of the black hole with and without the limiting
zero Hawking radiation. From the viewpoints of the constant
amplitude and decaying amplitude Hawking radiating configuration
with and without the logarithmic quantum gravity correction, the
present intrinsic geometric investigation shows that the nature of
the stability structure is very illuminative for the other
possible Hawking radiating black holes, and thus, our analysis
offers a very prominent tool for understanding the limiting
thermodynamic (un)stability structure, specifically for the most
puzzling Schwarzschild black hole, since its discovery.

Applicative analysis is further in progress to validate the
results with the conventional black holes arising from the modern
gauge field theories, higher dimensional gravity and gravitational
aspect of the (super)string theories. It is not surprising that
the present perspective provides the limiting intrinsic geometric
front to the stability structure of arbitrary finite parameter
Hawking radiating black holes and their possible generalizations.
Herewith, it may be anticipated that the nature of background
disturbances, vacuum fluctuations, statistical fluctuations,
limiting thermodynamic fluctuations and other perspective
fluctuations can be suitably modeled in the diverse possible
Hawking radiating black hole configurations. It follows from our
intrinsic geometric method, that the present analysis offers
perspective stability grounds, when applied to either the large
mass or Planck size black holes. It is expected further that the
present investigation would be an important factor in an
appropriate understanding of the statistical indicators under the
fluctuations of parameters, radiation frequency, quantum geometric
invariants and the other possible higher loop components arising
from the fluctuations of quantum gravity vacuum. Such explorations
are the matter of a future investigation.

\section*{Acknowledgement}
The work of S.B. is supported in part by the European Research
Council grant n.~226455, \textit{``SUPERSYMMETRY, QUANTUM GRAVITY
AND GAUGE FIELDS (SUPERFIELDS)''}. B.N.T. would like to thank
Professors V. Ravishankar, L. Bonora and S. Siwach for their
valuable support and encouragements; acknowledges the hospitality
of (i) Department of Physics, Banaras Hindu University, Varanasi,
India during the National Conference on New Trends in Field
Theories, (1st - 2nd Nov. 2008) and (ii) The Abdus Salam,
International Centre for Theoretical Physics Trieste, Italy during
the Spring School on Superstring Theory and Related Topics (22-30
March 2010), and the postdoctoral research fellowship of
\textit{``INFN, Italy''}, for the realization of this work.
\section*{Appendix-I: Coefficients in Quantum Corrections}
In this appendix, we provide explicit forms of the thermodynamic
correlation functions as the function of the radiation frequency
for the Hawking radiating configuration. Our analysis illustrates
that the physical properties of the specific correlations may
easily be exactly exploited in general. The definite behavior of
correlations as accounted in the concerned sections suggests that
the various intriguing constant and decaying amplitude examples of
radiating black hole solutions includes nice properties, i.e.,
that they do have a definite stability, except that the
determinant may be non-positive definite in some cases at specific
limit of the fluctuation and radiation frequency.

As mentioned in the main sections, these black hole configurations
are generically well-defined and indicate an interacting
statistical basis in general. We discover herewith that their
thermodynamic geometries indicate the possible nature of general
Hawking radiating black hole configurations. Significantly, we
notice from the very definition of intrinsic metric tensor that
the coefficients describing the thermodynamic correlations may be
enlisted as follows.
\subsection*{Appendix (A): Constant Amplitude Horizon Fluctuations}
Herewith, we provide the exact expressions for the constant
amplitude thermodynamic correlations with an arbitrary fluctuating
mass of the black hole. Explicitly, it turns out that the
functional nature of parameters within a small neighborhood of
statistical fluctuations introduced over an equilibrium ensemble
of configurations may precisely be divulged. Surprisingly, we can
expose in the framework of thermodynamic geometry that the
coefficients of the constant amplitude thermodynamic correlations
take the following exact and simple expressions.
\subsubsection*{Appendix (A1): Coefficients of the metric tensor}
\begin{scriptsize}
\begin{eqnarray}
g^{cMM}_{0}&:=& a L^2 u^4+4 a L^2 u^2+4 a L^2 \nonumber \\
g^{cMM}_{2}&:=& 144 a L^2 u^2 w^2-24 \pi u^2-2 \pi u^6-12 \pi u^4+32 a L^2 w^2 u^4-16 \pi+160 a L^2 w^2 \nonumber \\
g^{cMM}_{4}&:=& -640 \pi w^2+192 a L^2 u^4 w^4+2560 a L^2 w^4+1408 a L^2 u^2 w^4+48 \pi u^6 w^2+32 \pi w^2 u^4-448 \pi u^2 w^2 \nonumber \\
g^{cMM}_{6}&:=& 20480 a L^2 w^6-3584 \pi u^2 w^4-10240 \pi w^4+768 \pi u^4 w^4+3072 a L^2 u^2 w^6 \nonumber \\
g^{cMM}_{8}&:=& -20480 \pi u^2 w^6-8192 a L^2 u^2 w^8-2048 \pi w^6 u^4-81920 \pi w^6+81920 a L^2 w^8 \nonumber \\
g^{cMM}_{10}&:=& -327680 \pi w^8+131072 a L^2 w^{10}-65536 \pi u^2 w^8 \nonumber \\
g^{cMM}_{12}&:=&-524288 \pi w^{10} \nonumber \\
g^{cMw}_{0}&:=& -a L^2 u^2-2 a L^2 \nonumber \\
g^{cMw}_{2}&:=& -4 \pi u^4-16 \pi-8 a L^2 u^2 w^2-16 a L^2 w^2-16 \pi u^2 \nonumber \\
g^{cMw}_{4}&:=& 128 a L^2 w^4-128 \pi u^2 w^2-256 \pi w^2 \nonumber \\
g^{cMw}_{6}&:=& 1024 a L^2 w^6-1024 \pi w^4 \nonumber \\
g^{cww}_{0}&:=& -a L^2 u^2-2 a L^2 \nonumber \\
g^{cww}_{2}&:=& -2 \pi u^4-8 \pi+16 a L^2 w^2-8 \pi u^2 \nonumber \\
g^{cww}_{4}&:=& 128 \pi u^2 w^2+48 \pi w^2 u^4+64 \pi w^2+64 a L^2 u^2 w^4+640 a L^2 w^4 \nonumber \\
g^{cww}_{6}&:=& 3072 a L^2 w^6+2560 \pi w^4+1536 \pi u^2 w^4 \nonumber \\
g^{cww}_{8}&:=& 12288 \pi w^6
\end{eqnarray}
\end{scriptsize}
\subsubsection*{Appendix (A2): Coefficient of the determinant of the metric tensor}
\begin{scriptsize}
\begin{eqnarray}
g^{c}_{0}&:=&-a^2 L^4 u^4-4 a^2 L^4 u^2-4 a^2 L^4 \nonumber \\
g^{c}_{2}&:=&-16 a^2 L^4 u^2 w^2-64 a^2 L^4 w^2+8 a^2 L^4 w^2 u^4 \nonumber \\
g^{c}_{4}&:=&320 a^2 L^4 u^4 w^4+192 \pi u^6 a L^2 w^2+96 \pi^2 u^4+4 \pi^2 u^8+128 \pi^2 u^2 \nonumber \\&&
+64 \pi^2+768 \pi u^4 a L^2 w^2+32 \pi^2 u^6+1280 a^2 L^4 w^4+1024 a^2 L^4 u^2 w^4 \nonumber \\&&
+768 a L^2 u^2 \pi w^2 \nonumber \\
g^{c}_{6}&:=&1024 \pi^2 w^2+11264 \pi u^4 a L^2 w^4+1536 \pi u^6 a L^2 w^4+1536 a^2 L^4 w^6 u^4 \nonumber \\&&
+3840 \pi^2 u^2 w^2+1856 \pi^2 u^6 w^2+288 \pi^2 u^8 w^2+10240 a^2 L^4 u^2 w^6 \nonumber \\&&
+40960 a^2 L^4 w^6+16384 \pi a L^2 u^2 w^4+4224 \pi^2 u^4 w^2 \nonumber \\
g^{c}_{8}&:=&24576 \pi^2 u^2 w^4-20480 \pi^2 w^4+39936 \pi^2 u^4 w^4+16384 a^2 L^4 u^2 w^8  \nonumber \\&&
+32768 \pi u^4 a L^2 w^6+409600 a^2 L^4 w^8+11264 \pi^2 u^6 w^4+98304 a L^2 u^2 \pi w^6 \nonumber \\
g^{c}_{10}&:=&-65536 a^2 L^4 u^2 w^{10}+1835008 a^2 L^4 w^{10}-655360 \pi^2 w^6+24576 \pi^2 u^4 w^6  \nonumber \\&&
-65536 a L^2 u^4 w^8 \pi-12288 \pi^2 u^6 w^6-229376 \pi^2 u^2 w^6  \nonumber \\
g^{c}_{12}&:=&-3145728 \pi^2 u^2 w^8-6553600 \pi^2 w^8+3145728 a^2 L^4 w^{12}  \nonumber \\&&
-589824 \pi^2 w^8 u^4-1048576 \pi w^{10} a L^2 u^2 \nonumber \\
g^{c}_{14}&:=&-29360128 \pi^2 w^{10}-9437184 \pi^2 w^{10} u^2 \nonumber \\
g^{c}_{16}&:=&-50331648 \pi^2 w^{12}
\end{eqnarray}
\end{scriptsize}
\subsubsection*{Appendix (A3): Coefficients of the scalar curvature}
\begin{scriptsize}
\begin{eqnarray}
r^{c}_{0}&:=& -8 a^3 L^6 u^6-24 a^3 L^6 u^4-16 a^3 L^6-a^3 L^6 u^8-32 a^3 L^6 u^2  \nonumber \\
r^{c}_{2}&:=& -6 a^2 L^4 \pi u^{10}-60 a^2 L^4 \pi u^8-256 a^3 w^2 L^6 u^6-1024 a^3 w^2 L^6 u^2  \nonumber \\&&
-480 a^2 L^4 u^4 \pi-32 a^3 w^2 L^6 u^8-512 a^3 w^2 L^6-192 a^2 \pi L^4-240 a^2 L^4 \pi u^6  \nonumber \\&&
-480 a^2 L^4 \pi u^2-768 a^3 w^2 L^6 u^4  \nonumber \\
r^{c}_{4}&:=& -14336 a^3 w^4 L^6 u^2-12032 a^3 w^4 L^6 u^4-256 a \pi^2 L^2-2112 a^2 w^2 L^4 \pi u^8  \nonumber \\&&
-15360 a^2 w^2 \pi L^4-11136 a^2 w^2 L^4 \pi u^6-48 a L^2 \pi^2 u^{10}-6144 a^3 w^4 L^6  \nonumber \\&&
-144 a^2 w^2 L^4 \pi u^{10}-640 a L^2 \pi^2 u^6-768 a L^2 \pi^2 u^2-4352 a^3 w^4 L^6 u^6  \nonumber \\&&
-4 a L^2 \pi^2 u^{12}-960 a L^2 \pi^2 u^4-33024 a^2 w^2 L^4 \pi u^2-576 a^3 w^4 L^6 u^8  \nonumber \\&&
-240 a L^2 \pi^2 u^8-27648 a^2 w^2 L^4 u^4 \pi  \nonumber \\
r^{c}_{6}&:=& 1344 \pi^3 u^10+8960 \pi^3 u^6+10752 \pi^3 u^4+224 \pi^3 u^12+4480 \pi^3 u^8  \nonumber \\&&
-3584 a w^2 L^2 \pi^2 u^{10}-1275904 a^2 w^4 L^4 \pi u^2-256 a w^2 L^2 \pi^2 u^{12}  \nonumber \\&&
-364032 a^2 w^4 L^4 \pi u^6-62720 a^2 w^4 L^4 \pi u^8-986112 a^2 w^4 L^4 u^4 \pi  \nonumber \\&&
-20480 a w^2 L^2 \pi^2 u^8-61440 a w^2 L^2 \pi^2 u^6+16 \pi^3 u^{14}+7168 u^2 \pi^3  \nonumber \\&&
+2048 \pi^3-32768 a w^2 \pi^2 L^2-638976 a^2 w^4 \pi L^4-77824 a^3 w^6 L^6 u^6  \nonumber \\&&
-8192 a^3 w^6 L^6 u^8-278528 a^3 w^6 L^6 u^4-458752 a^3 w^6 L^6 u^2  \nonumber \\&&
-102400 a w^2 L^2 \pi^2 u^4-90112 a w^2 L^2 \pi^2 u^2-3840 a^2 w^4 L^4 \pi u^{10}  \nonumber \\&&
-294912 a^3 w^6 L^6  \nonumber \\
r^{c}_{8}&:=& 114688 w^2 \pi^3-185344 a w^4 L^2 \pi^2 u^{10}-8646656 a^2 w^6 L^4 \pi u^6  \nonumber \\&&
-980992 a w^4 L^2 \pi^2 u^8-4579328 a w^4 L^2 \pi^2 u^4-2809856 a w^4 L^2 \pi^2 u^6  \nonumber \\&&
-67584 a^2 w^6 L^4 \pi u^{10}-34226176 a^2 w^6 L^4 \pi u^2-4014080 a w^4 L^2 \pi^2 u^2  \nonumber \\&&
-14848 a w^4 L^2 \pi^2 u^{12}-1343488 a^2 w^6 L^4 \pi u^8-25100288 a^2 w^6 L^4 u^4 \pi  \nonumber \\&&
-17891328 a^2 w^6 \pi L^4-77824 a^3 w^8 L^6 u^8-16515072 a^3 w^8 L^6 u^2  \nonumber \\&&
-1228800 a^3 w^8 L^6 u^6-1474560 a w^4 \pi^2 L^2-6815744 a^3 w^8 L^6 u^4  \nonumber \\&&
+352256 w^2 u^2 \pi^3+128 w^2 \pi^3 u^{14}+454656 w^2 \pi^3 u^4+3328 w^2 \pi^3 u^{12}  \nonumber \\&&
+317440 w^2 \pi^3 u^6+29184 w^2 \pi^3 u^{10}+128000 w^2 \pi^3 u^8-14352384 a^3 w^8 L^6  \nonumber \\
r^{c}_{10}&:=& 3538944 w^4 \pi^3-20709376 a w^6 L^2 \pi^2 u^8-424148992 a^2 w^8 L^4 u^4 \pi  \nonumber \\&&
-637272064 a^2 w^8 L^4 \pi u^2-94896128 a w^6 L^2 \pi^2 u^2-107741184 a w^6 L^2 \pi^2 u^4  \nonumber \\&&
-126976000 a^2 w^8 L^4 \pi u^6-15876096 a^2 w^8 L^4 \pi u^8-516096 a^2 w^8 L^4 \pi u^{10}  \nonumber \\&&
-34603008 a w^6 \pi^2 L^2-99614720 a^3 w^{10} L^6 u^4-12058624 a^3 w^{10} L^6 u^6  \nonumber \\&&
-393216 a^3 w^{10} L^6 u^8-322961408 a^3 w^{10} L^6 u^2-356253696 a^2 w^8 \pi L^4  \nonumber \\&&
-64094208 a w^6 L^2 \pi^2 u^6-3342336 a w^6 L^2 \pi^2 u^{10}-196608 a w^6 L^2 \pi^2 u^{12}  \nonumber \\&&
-9216 w^4 \pi^3 u^{14}-86016 w^4 \pi^3 u^{12}+9043968 w^4 u^2 \pi^3-196608 w^4 \pi^3 u^{10}  \nonumber \\&&
+9043968 w^4 \pi^3 u^4+614400 w^4 \pi^3 u^8+4177920 w^4 \pi^3 u^6-339738624 a^3 w^{10} L^6  \nonumber \\
r^{c}_{12}&:=& 97517568 w^6 \pi^3-1348468736 a w^8 L^2 \pi^2 u^2-1376256 a^2 w^{10} L^4 \pi u^{10}  \nonumber \\&&
-1099694080 a^2 w^{10} L^4 \pi u^6-24444928 a w^8 L^2 \pi^2 u^{10}-4592762880 a^2 w^{10} L^4 u^4 \pi  \nonumber \\&&
-8088715264 a^2 w^{10} L^4 \pi u^2-97779712 a^2 w^{10} L^4 \pi u^8-5058330624 a^2 w^{10} \pi L^4  \nonumber \\&&
-832569344 a^3 w^{12} L^6 u^4-66060288 a^3 w^{12} L^6 u^6-3640655872 a^3 w^{12} L^6 u^2  \nonumber \\&&
-786432 a^3 w^{12} L^6 u^8-493879296 a w^8 \pi^2 L^2-209977344 a w^8 L^2 \pi^2 u^8  \nonumber \\&&
-737280 a w^8 L^2 \pi^2 u^{12}-783810560 a w^8 L^2 \pi^2 u^6-1462239232 a w^8 L^2 \pi^2 u^4  \nonumber \\&&
-4580179968 a^3 w^{12} L^6-73728 w^6 \pi^3 u^{14}+211812352 w^6 u^2 \pi^3+166723584 w^6 \pi^3 u^4  \nonumber \\&&
-1409024 w^6 \pi^3 u^{12}+46530560 w^6 \pi^3 u^6-7274496 w^6 \pi^3 u^{10}-7536640 w^6 \pi^3 u^8  \nonumber \\
r^{c}_{14}&:=& -37849399296 a^3 w^{14} L^6-69172461568 a^2 w^{12} L^4 \pi u^2-50734301184 a^2 w^{12} \pi L^4  \nonumber \\&&
-5200936960 a w^{10} L^2 \pi^2 u^6-1015021568 a w^{10} L^2 \pi^2 u^8-12213813248 a w^{10} L^2 \pi^2 u^2  \nonumber \\&&
+3334471680 w^8 \pi^3 u^4-3801088 w^8 \pi^3 u^{12}-5429526528 a^2 w^{12} L^4 \pi u^6+969932800 w^8 \pi^3 u^6  \nonumber \\&&
-65011712 a w^{10} L^2 \pi^2 u^{10}-32243712 w^8 \pi^3 u^{10}-24964497408 a^3 w^{14} L^6 u^2  \nonumber \\&&
+2365587456 w^8 \pi^3+4638900224 w^8 u^2 \pi^3+34078720 w^8 \pi^3 u^8-4831838208 a w^{10} \pi^2 L^2  \nonumber \\&&
-288358400 a^2 w^{12} L^4 \pi u^8-184549376 a^3 w^{14} L^6 u^6-11760828416 a w^{10} L^2 \pi^2 u^4  \nonumber \\&&
-31314673664 a^2 w^{12} L^4 u^4 \pi-3959422976 a^3 w^{14} L^6 u^4  \nonumber \\
r^{c}_{16}&:=& 42278584320 w^{10} \pi^3-39460012032 a w^{12} \pi^2 L^2-190857609216 a^3 w^{16} L^6  \nonumber \\&&
+144703488 w^{10} \pi^3 u^{10}-201326592 a^3 w^{16} L^6 u^6+74692165632 w^{10} u^2 \pi^3  \nonumber \\&&
+2076180480 w^{10} \pi^3 u^8+49425678336 w^{10} \pi^3 u^4+9437184 w^{10} \pi^3 u^{12}  \nonumber \\&&
+14973665280 w^{10} \pi^3 u^6-301989888 a^2 w^{14} L^4 \pi u^8-103616086016 a^3 w^{16} L^6 u^2  \nonumber \\&&
-390439370752 a^2 w^{14} L^4 \pi u^2-54358179840 a w^{12} L^2 \pi^2 u^4  \nonumber \\&&
-1820327936 a w^{12} L^2 \pi^2 u^8-12582912 a w^{12} L^2 \pi^2 u^{10}  \nonumber \\&&
-13774094336 a^2 w^{14} L^4 \pi u^6-129117454336 a^2 w^{14} L^4 u^4 \pi  \nonumber \\&&
-352724189184 a^2 w^{14} \pi L^4-74222403584 a w^{12} L^2 \pi^2 u^2  \nonumber \\&&
-9797894144 a^3 w^{16} L^6 u^4-16944988160 a w^{12} L^2 \pi^2 u^6  \nonumber \\
r^{c}_{18}&:=& 905969664 w^{12} \pi^3 u^10+121131499520 w^{12} \pi^3 u^6-343597383680 a w^{14} L^2 \pi^2 u^2  \nonumber \\&&
-1384590082048 a^2 w^{16} L^4 \pi u^2+15602810880 w^{12} \pi^3 u^8+796716433408 w^{12} u^2 \pi^3  \nonumber \\&&
+455803404288 w^{12} \pi^3 u^4+518617300992 w^{12} \pi^3-1633161314304 a^2 w^{16} \pi L^4  \nonumber \\&&
-240518168576 a^3 w^{18} L^6 u^2+536870912 a w^{14} L^2 \pi^2 u^8-129922760704 a w^{14} L^2 \pi^2 u^4  \nonumber \\&&
-15569256448 a w^{14} L^2 \pi^2 u^6-288299679744 a^2 w^{16} L^4 u^4 \pi-8589934592 a^3 w^{18} L^6 u^4  \nonumber \\&&
-335007449088 a w^{14} \pi^2 L^2-12213813248 a^2 w^{16} L^4 \pi u^6-523986010112 a^3 w^{18} L^6  \nonumber \\
r^{c}_{20}&:=& 2486786064384 w^{14} \pi^3 u^4-377957122048 a^3 w^{20} L^6-4587025072128 a^2 w^{18} \pi L^4  \nonumber \\&&
+4294967296 a^3 w^{20} L^6 u^4-240518168576 a^2 w^{18} L^4 u^4 \pi-1541893259264 a w^{16} L^2 \pi^2 u^2  \nonumber \\&&
-2735894167552 a^2 w^{18} L^4 \pi u^2+4303557230592 w^{14} \pi^3+805306368 a w^{16} L^2 \pi^2 u^8  \nonumber \\&&
+5471788335104 w^{14} u^2 \pi^3-2744484102144 a w^{16} \pi^2 L^2-153545080832 a w^{16} L^2 \pi^2 u^4  \nonumber \\&&
+36238786560 w^{14} \pi^3 u^8+25769803776 a w^{16} L^2 \pi^2 u^6-240518168576 a^3 w^{20} L^6 u^2  \nonumber \\&&
+485868175360 w^{14} \pi^3 u^6+5368709120 a^2 w^{18} L^4 \pi u^6  \nonumber \\
r^{c}_{22}&:=& -6734508720128 a w^{18} L^2 \pi^2 u^2+773094113280 w^{16} \pi^3 u^6  \nonumber \\&&
-412316860416 a w^{18} L^2 \pi^2 u^4-5772436045824 a^2 w^{20} \pi L^4  \nonumber \\&&
-2027224563712 a^2 w^{20} L^4 \pi u^2+85899345920 a^2 w^{20} L^4 u^4 \pi  \nonumber \\&&
-17042430230528 a w^{18} \pi^2 L^2+7395933683712 w^{16} \pi^3 u^4  \nonumber \\&&
+23811298689024 w^{16} \pi^3+1649267441664 a^3 L^6 w^{22}+23347442221056 w^{16} u^2 \pi^3  \nonumber \\
r^{c}_{24}&:=& 3298534883328 a^3 L^6 w^{24}+56487409876992 w^{18} u^2 \pi^3  \nonumber \\&&
-1236950581248 a w^{20} L^2 \pi^2 u^4+824633720832 a^2 w^{22} L^4 \pi u^2  \nonumber \\&&
-68719476736000 a w^{20} \pi^2 L^2+9277129359360 w^{18} \pi^3 u^4  \nonumber \\&&
+84387517431808 w^{18} \pi^3+3298534883328 a^2 w^{22} \pi L^4  \nonumber \\&&
-20066087206912 a w^{20} L^2 \pi^2 u^2  \nonumber \\
r^{c}_{26}&:=& -26388279066624 a w^{22} L^2 \pi^2 u^2+173722837188608 w^{20} \pi^3  \nonumber \\&&
+13194139533312 a^2 \pi L^4 w^{24}+59373627899904 w^{20} u^2 \pi^3  \nonumber \\&&
-158329674399744 a w^{22} \pi^2 L^2  \nonumber \\
r^{c}_{28}&:=& 158329674399744 \pi^3 w^{22}-158329674399744 a \pi^2 L^2 w^{24}
\end{eqnarray}
\end{scriptsize}
\subsection*{Appendix (B): Decaying Amplitude Horizon Fluctuations}
In this case, we may again explicitly provide the exact
expressions for the decaying amplitude thermodynamic correlations
with an arbitrary fluctuating mass of the Hawking radiating black
hole. It turns out that the functional nature of parameters,
within a small neighborhood of statistical fluctuations introduced
over an equilibrium ensemble of configurations, may precisely be
divulged as in the case of the constant amplitude thermodynamic
fluctuations. Similarly, we can expose in this framework that the
coefficients of the decaying amplitude thermodynamic correlations
take the following exact and simple expressions.
\subsubsection*{Appendix (B1): Coefficients of the metric tensor}
\begin{scriptsize}
\begin{eqnarray}
g^{dMM}_0&:=& a^3 m^4 L^2+16 a^5 m^6 L^2 w^2-\pi a^4 m^4-16 \pi a^6 w^2 m^6   \nonumber \\
g^{dMM}_2&:=& a m^2 L^2+2 \pi a^2 m^2+112 a^3 m^4 L^2 w^2+768 \pi a^6 w^4 m^6-16 \pi w^2 a^4 m^4 \nonumber \\
g^{dMM}_4&:=& -4096 a^5 m^6 L^2 w^6+4864 a^3 m^4 L^2 w^4-\pi+80 a m^2 L^2 w^2-1792 \pi w^4 a^4 m^4+112 \pi w^2 a^2 m^2 \nonumber \\
g^{dMM}_6&:=& 3328 \pi w^4 a^2 m^2-80 \pi w^2+86016 a^3 m^4 L^2 w^6+2560 a m^2 L^2 w^4-28672 \pi w^6 a^4 m^4 \nonumber \\
g^{dMM}_8&:=& 40960 a m^2 L^2 w^6+524288 a^3 m^4 L^2 w^8-2560 \pi w^4+53248 \pi w^6 a^2 m^2 \nonumber \\
g^{dMM}_{10}&:=& 327680 \pi w^8 a^2 m^2+327680 a m^2 L^2 w^8-40960 \pi w^6 \nonumber \\
g^{dMM}_{12}&:=&-327680 \pi w^8+1048576 a m^2 L^2 w^{10} \nonumber \\
g^{dMM}_{14}&:=&-1048576 \pi w^{10}      \nonumber \\
g^{dMw}_{0}&:=&a^3 m^4 L^2-\pi a^4 m^4 \nonumber \\
g^{dMw}_{2}&:=&16 \pi w^2 a^4 m^4+2 \pi a^2 m^2+16 a^3 m^4 L^2 w^2 \nonumber \\
g^{dMw}_{4}&:=&-\pi+32 a m^2 L^2 w^2  \nonumber \\
g^{dMw}_{6}&:=&-16 \pi w^2-512 \pi w^4 a^2 m^2+1024 a m^2 L^2 w^4  \nonumber \\
g^{dMw}_{8}&:=&256 \pi w^4+8192 a m^2 L^2 w^6  \nonumber \\
g^{dMw}_{10}&:=&4096 \pi w^6 \nonumber \\
g^{dww}_{0}&:=& a^3 m^4 L^2-\pi a^4 m^4 \nonumber \\
g^{dww}_{2}&:=& -a m^2 L^2+48 \pi w^2 a^4 m^4+2 \pi a^2 m^2  \nonumber \\
g^{dww}_{4}&:=& -\pi-256 a^3 m^4 L^2 w^4+16 a m^2 L^2 w^2-64 \pi w^2 a^2 m^2 \nonumber \\
g^{dww}_{6}&:=& 1280 a m^2 L^2 w^4+16 \pi w^2-1536 \pi w^4 a^2 m^2  \nonumber \\
g^{dww}_{8}&:=& 1280 \pi w^4+12288 a m^2 L^2 w^6  \nonumber \\
g^{dww}_{10}&:=&12288 \pi M^10 w^6
\end{eqnarray}
\end{scriptsize}
\subsubsection*{Appendix (B2): Coefficients of the determinant of the metric tensor}
\begin{scriptsize}
\begin{eqnarray}
g^{d}_{0}&:=&-2 a^5 m^6 L^2 \pi+\pi^2 a^6 m^6+a^4 m^6 L^4-48 a^8 m^8 \pi^2 w^2+96 a^7 m^8 \pi w^2 L^2-48 a^6 m^8 L^4 w^2 \nonumber \\
g^{d}_{2}&:=&(-3 a^4 m^4 \pi^2+a^2 m^4 L^4+2 a^3 m^4 L^2 \pi+128 a^6 m^6 \pi^2 w^2+32 a^4 m^6 L^4 w^2  \nonumber \\ &&
+1280 a^8 m^8 \pi^2 w^4-160 a^5 m^6 L^2 w^2 \pi-1280 a^6 m^8 w^4 L^4)  \nonumber \\
g^{d}_{4}&:=&2560 a^5 m^6 L^2 w^4 \pi-2560 a^4 m^6 L^4 w^4+3 a^2 m^2 \pi^2-2816 a^6 m^6 \pi^2 w^4-144 \pi^2 w^2 a^4 m^4  \nonumber \\ &&
+32 a^2 m^4 L^4 w^2+96 a^3 m^4 L^2 w^2 \pi-4096 a^6 m^8 L^4 w^6-24576 a^7 m^8 L^2 w^6 \pi)  \nonumber \\
g^{d}_{6}&:=&(96 a^2 m^2 \pi^2 w^2+3840 \pi^2 w^4 a^4 m^4-1280 a^2 m^4 L^4 w^4-73728 a^6 m^6 \pi^2 w^6  \nonumber \\ &&
+65536 a^6 m^8 w^8 L^4-3072 \pi w^4 a^3 m^4 L^2+172032 a^5 m^6 L^2 w^6 \pi-131072 a^4 m^6 L^4 w^6-\pi^2)  \nonumber \\
g^{d}_{8}&:=&(-81920 a^2 m^4 L^4 w^6-212992 \pi w^6 a^3 m^4 L^2-3584 a^2 m^2 \pi^2 w^4  \nonumber \\ &&
+184320 a^4 m^4 \pi^2 w^6-2031616 a^4 m^6 L^4 w^8-32 \pi^2 w^2+1572864 a^5 m^6 L^2 w^8 \pi)  \nonumber \\
g^{d}_{10}&:=&(-3538944 a^3 m^4 L^2 w^8 \pi-10485760 a^4 m^6 L^4 w^10-1638400 a^2 m^4 L^4 w^8  \nonumber \\&&
+1280 \pi^2 w^4-196608 a^2 m^2 \pi^2 w^6+1572864 a^4 m^4 \pi^2 w^8)  \nonumber \\
g^{d}_{12}&:=&(-14680064 a^2 m^4 L^4 w^{10}-2949120 a^2 m^2 \pi^2 w^8-18874368 a^3 m^4 L^2 w^{10} \pi+81920 \pi^2 w^6)  \nonumber \\
g^{d}_{14}&:=&(1638400 \pi^2 w^8-14680064 a^2 m^2 \pi^2 w^{10}-50331648 a^2 m^4 L^4 w^{12})  \nonumber \\
g^{d}_{16}&:=&14680064 \pi^2 w^{10}  \nonumber \\
g^{d}_{18}&:=& 50331648 \pi^2 w^{12}
\end{eqnarray}
\end{scriptsize}
\subsubsection*{Appendix (B3): Coefficients of the scalar curvature}
\begin{scriptsize}
\begin{eqnarray}
r^{d}_{0}&:=&3 a^{11} \pi^2 m^{12} L^2-3 a^{10} \pi m^{12} L^4-a^{12} \pi^3 m^{12}+a^9 m^{12} L^6  \nonumber \\
r^{d}_{2}&:=& 128 a^{12} \pi^3 m^{12} w^2+6 a^{10} \pi^3 m^{10}+10 a^8 \pi m^{10} L^4-14 a^9 \pi^2 m^{10} L^2  \nonumber \\ &&
-288 a^{11} \pi^2 m^{12} L^2 w^2+5120 a^{13} \pi^2 m^{14} L^2 w^4-2048 a^{14} \pi^3 m^{14} w^4  \nonumber \\ &&
+1024 a^{11} m^{14} L^6 w^4-4096 a^{12} \pi m^{14} L^4 w^4-32 a^9 m^{12} L^6 w^2+192 a^{10} \pi m^{12} L^4 w^2  \nonumber \\ &&
-2 a^7 m^{10} L^6  \nonumber \\
r^{d}_{4}&:=& -a^5 m^8 L^6-40704 a^{11} \pi^2 m^{12} L^2 w^4+13056 a^{12} \pi^3 m^{12} w^4-384 a^8 \pi m^{10} L^4 w^2  \nonumber \\ &&
+64 a^7 m^{10} L^6 w^2+65536 a^{14} \pi^3 m^{14} w^6+992 a^9 \pi^2 m^{10} L^2 w^2-6 a^6 \pi m^8 L^4  \nonumber \\ &&
-81920 a^{12} \pi m^{14} L^4 w^6-32768 a^{13} \pi^2 m^{14} L^2 w^6-15 a^8 \pi^3 m^8-12800 a^9 m^{12} L^6 w^4  \nonumber \\ &&
+49152 a^{11} m^{14} L^6 w^6-672 a^{10} \pi^3 m^{10} w^2+22 a^7 \pi^2 m^8 L^2+40448 a^{10} \pi m^{12} L^4 w^4  \nonumber \\
r^{d}_{6}&:=& 1572864 a^{14} \pi^3 m^{14} w^8+2097152 a^{12} \pi m^{14} L^4 w^8-360448 a^{12} \pi^3 m^{12} w^6  \nonumber \\ &&
+26112 a^7 m^{10} L^6 w^4-1280 a^7 \pi^2 m^8 L^2 w^2-4194304 a^{13} \pi^2 m^{14} L^2 w^8  \nonumber \\ &&
-12 a^5 \pi^2 m^6 L^2-102400 a^8 \pi m^{10} L^4 w^4+32 a^5 m^8 L^6 w^2-2 a^4 \pi m^6 L^4  \nonumber \\ &&
-47616 a^{10} \pi^3 m^{10} w^4+524288 a^{11} m^{14} L^6 w^8+128512 a^9 \pi^2 m^{10} L^2 w^4  \nonumber \\ &&
+20 a^6 \pi^3 m^6+1081344 a^{10} \pi m^{12} L^4 w^6+1440 a^8 \pi^3 m^8 w^2+96 a^6 \pi m^8 L^4 w^2  \nonumber \\ &&
-49152 a^{11} \pi^2 m^{12} L^2 w^6-647168 a^9 m^{12} L^6 w^6  \nonumber \\
r^{d}_{8}&:=& -15 a^4 \pi^3 m^4+960 a^5 \pi^2 m^6 L^2 w^2+909312 a^{10} \pi^3 m^{10} w^6+45416448 a^{11} \pi^2 m^{12} L^2 w^8  \nonumber \\ &&
+58720256 a^{12} \pi m^{14} L^4 w^{10}-13893632 a^{10} \pi m^{12} L^4 w^8+1310720 a^7 m^{10} L^6 w^6+110080 a^8 \pi^3 m^8 w^4  \nonumber \\ &&
+64 a^4 \pi m^6 L^4 w^2+3840 a^5 m^8 L^6 w^4+a^2 \pi m^4 L^4+876544 a^9 \pi^2 m^{10} L^2 w^6-96 a^3 m^6 L^6 w^2  \nonumber \\ &&
-8388608 a^{11} m^{14} L^6 w^{10}-a^3 \pi^2 m^4 L^2-3096576 a^8 \pi m^{10} L^4 w^6-186368 a^7 \pi^2 m^8 L^2 w^4  \nonumber \\ &&
-18546688 a^{12} \pi^3 m^{12} w^8+64000 a^6 \pi m^8 L^4 w^4-1600 a^6 \pi^3 m^6 w^2-11010048 a^9 m^{12} L^6 w^8  \nonumber \\ &&
-25165824 a^{13} \pi^2 m^{14} L^2 w^{10}  \nonumber \\
r^{d}_{10}&:=& 268435456 a^{12} \pi m^{14} L^4 w^{12}+59506688 a^8 \pi m^{10} L^4 w^8-150994944 a^{12} \pi^3 m^{12} w^{10}  \nonumber \\ &&
+125952 a^5 \pi^2 m^6 L^2 w^4-960495616 a^{10} \pi m^{12} L^4 w^{10}+2 a \pi^2 m^2 L^2+79822848 a^{10} \pi^3 m^{10} w^8  \nonumber \\ &&
+12320768 a^7 m^{10} L^6 w^8+960 a^4 \pi^3 m^4 w^2-7680 a^3 m^6 L^6 w^4+11776 a^4 \pi m^6 L^4 w^4  \nonumber \\ &&
-201326592 a^{11} m^{14} L^6 w^{12}+937426944 a^{11} \pi^2 m^{12} L^2 w^{10}-32 a^2 \pi m^4 L^4 w^2  \nonumber \\ &&
+48234496 a^9 m^{12} L^6 w^{10}-166592512 a^9 \pi^2 m^{10} L^2 w^8+2392064 a^6 \pi m^8 L^4 w^6  \nonumber \\&&
+6 a^2 \pi^3 m^2-942080 a^8 \pi^3 m^8 w^6-163840 a^5 m^8 L^6 w^6+201326592 a^{13} \pi^2 m^{14} L^2 w^{12}  \nonumber \\ &&
-608 a^3 \pi^2 m^4 L^2 w^2-153600 a^6 \pi^3 m^6 w^4-1998848 a^7 \pi^2 m^8 L^2 w^6  \nonumber \\
r^{d}_{12}&:=& 425984 a^4 \pi m^6 L^4 w^6-1073741824 a^{12} \pi m^{14} L^4 w^{14}+224 a \pi^2 m^2 L^2 w^2  \nonumber \\ &&
-21102592 a^5 m^8 L^6 w^8-931135488 a^7 m^{10} L^6 w^{10}+5486149632 a^{11} \pi^2 m^{12} L^2 w^{12}  \nonumber \\ &&
-37120 a^3 \pi^2 m^4 L^2 w^4-327680 a^6 \pi^3 m^6 w^6-1073741824 a^{11} m^{14} L^6 w^{14}  \nonumber \\ &&
-14596177920 a^{10} \pi m^{12} L^4 w^{12}-180879360 a^8 \pi^3 m^8 w^8-9216 a^2 \pi m^4 L^4 w^4  \nonumber \\ &&
-56360960 a^6 \pi m^8 L^4 w^8-5475663872 a^9 \pi^2 m^{10} L^2 w^{10}+282329088 a^7 \pi^2 m^8 L^2 w^8  \nonumber \\ &&
+123648 a^4 \pi^3 m^4 w^4+4328521728 a^9 m^{12} L^6 w^{12}-288 a^2 \pi^3 m^2 w^2-73728 a^3 m^6 L^6 w^6  \nonumber \\ &&
+150994944 a^{12} \pi^3 m^{12} w^{12}+1277165568 a^{10} \pi^3 m^{10} w^{10}+1179648 a^5 \pi^2 m^6 L^2 w^6  \nonumber \\ &&
-\pi^3+5054136320 a^8 \pi m^{10} L^4 w^{10}  \nonumber \\
r^{d}_{14}&:=& -38184943616 a^7 m^{10} L^6 w^{12}+61740154880 a^9 m^{12} L^6 w^{14}+4608 a \pi^2 m^2 L^2 w^4  \nonumber \\ &&
+115695681536 a^8 \pi m^{10} L^4 w^{12}-792723456 a^5 m^8 L^6 w^{10}-319488 a^2 \pi m^4 L^4 w^6  \nonumber \\ &&
-4330618880 a^8 \pi^3 m^8 w^{10}+14417920 a^4 \pi m^6 L^4 w^8+458752 a^3 \pi^2 m^4 L^2 w^6  \nonumber \\ &&
+32 \pi^3 w^2-68954357760 a^9 \pi^2 m^{10} L^2 w^{12}-85899345920 a^{10} \pi m^{12} L^4 w^{14}  \nonumber \\ &&
+12322865152 a^7 \pi^2 m^8 L^2 w^{10}+1572864 a^4 \pi^3 m^4 w^6+12976128 a^3 m^6 L^6 w^8  \nonumber \\ &&
+230686720 a^6 \pi^3 m^6 w^8-52736 a^2 \pi^3 m^2 w^4-5968494592 a^6 \pi m^8 L^4 w^{10}  \nonumber \\ &&
+5133828096 a^{10} \pi^3 m^{10} w^12-246939648 a^5 \pi^2 m^6 L^2 w^8  \nonumber \\
r^{d}_{16}&:=& 117309440 a^3 \pi^2 m^4 L^2 w^8-177905598464 a^6 \pi m^8 L^4 w^{12}+6094848 a^2 \pi m^4 L^4 w^8  \nonumber \\ &&
-38084280320 a^8 \pi^3 m^8 w^{12}+9216 \pi^3 w^4-678604832768 a^7 m^{10} L^6 w^{14}-15737028608 a^5 m^8 L^6 w^{12}  \nonumber \\ &&
+373662154752 a^9 m^{12} L^6 w^{16}+223472517120 a^7 \pi^2 m^8 L^2 w^{12}-12255756288 a^5 \pi^2 m^6 L^2 w^{10}  \nonumber \\ &&
-355945414656 a^9 \pi^2 m^{10} L^2 w^{14}+763363328 a^4 \pi m^6 L^4 w^{10}+7675576320 a^6 \pi^3 m^6 w^{10}  \nonumber \\ &&
+1293858897920 a^8 \pi m^{10} L^4 w^{14}-1236992 a^2 \pi^3 m^2 w^6-141733920768 a^{10} \pi m^{12} L^4 w^{16}  \nonumber \\ &&
-160825344 a^4 \pi^3 m^4 w^8-14495514624 a^{10} \pi^3 m^{10} w^{14}-417792 a \pi^2 m^2 L^2 w^6  \nonumber \\ &&
+641728512 a^3 m^6 L^6 w^{10}  \nonumber \\
r^{d}_{18}&:=& -56371445760 a^8 \pi^3 m^8 w^{14}+824633720832 a^9 m^{12} L^6 w^{18}+717225984 a^2 \pi m^4 L^4 w^{10}  \nonumber \\ &&
+2037961981952 a^7 \pi^2 m^8 L^2 w^{14}-7423918080 a^4 \pi^3 m^4 w^{10}+54263808 a^2 \pi^3 m^2 w^8  \nonumber \\ &&
+319488 \pi^3 w^6+4961861632 a^3 \pi^2 m^4 L^2 w^{10}+12683575296 a^3 m^6 L^6 w^{12}-26738688 a \pi^2 m^2 L^2 w^8  \nonumber \\ &&
-6571299962880 a^7 m^{10} L^6 w^{16}+7189775253504 a^8 \pi m^{10} L^4 w^16-463856467968 a^9 \pi^2 m^{10} L^2 w^{16}  \nonumber \\ &&
-2780991324160 a^6 \pi m^8 L^4 w^{14}-271790899200 a^5 \pi^2 m^6 L^2 w^{12}-182536110080 a^5 m^8 L^6 w^{14}  \nonumber \\ &&
+28655484928 a^4 \pi m^6 L^4 w^{12}+104018739200 a^6 \pi^3 m^6 w^{12}  \nonumber \\
r^{d}_{20}&:=& -24592982736896 a^6 \pi m^8 L^4 w^{16}-3298534883328 a^5 \pi^2 m^6 L^2 w^{14}  \nonumber \\ &&
+579820584960 a^8 \pi^3 m^8 w^{16}+15668040695808 a^8 \pi m^{10} L^4 w^18+536870912000 a^6 \pi^3 m^6 w^{14}  \nonumber \\ &&
-490733568 a \pi^2 m^2 L^2 w^{10}-33809982554112 a^7 m^{10} L^6 w^{18}+104522055680 a^3 \pi^2 m^4 L^2 w^{12}  \nonumber \\ &&
+8684423872512 a^7 \pi^2 m^8 L^2 w^{16}+3670016000 a^2 \pi^3 m^2 w^{10}-6094848 \pi^3 w^8  \nonumber \\ &&
+67645734912 a^3 m^6 L^6 w^{14}+22548578304 a^2 \pi m^4 L^4 w^{12}+627065225216 a^4 \pi m^6 L^4 w^{14}  \nonumber \\ &&
-1224065679360 a^5 m^8 L^6 w^{16}-138412032000 a^4 \pi^3 m^4 w^{12}  \nonumber \\
r^{d}_{22}&:=& -116135915683840 a^6 \pi m^8 L^4 w^{18}+11544872091648 a^7 \pi^2 m^8 L^2 w^{18}  \nonumber \\ &&
+8134668058624 a^4 \pi m^6 L^4 w^{16}-773094113280 a^6 \pi^3 m^6 w^{16}+7046430720 a \pi^2 m^2 L^2 w^{12}  \nonumber \\ &&
+370440929280 a^2 \pi m^4 L^4 w^{14}+89724551168 a^2 \pi^3 m^2 w^{12}-717225984 \pi^3 w^{10}  \nonumber \\ &&
-1211180777472 a^4 \pi^3 m^4 w^{14}-21698174779392 a^5 \pi^2 m^6 L^2 w^{16}-4260607557632 a^5 m^8 L^6 w^{18}  \nonumber \\ &&
-72567767433216 a^7 m^{10} L^6 w^{20}+1170378588160 a^3 \pi^2 m^4 L^2 w^{14}-2010044694528 a^3 m^6 L^6 w^{16}  \nonumber \\
r^{d}_{24}&:=& -22548578304 \pi^3 w^{12}-226499395321856 a^6 \pi m^8 L^4 w^{20}+537944653824 a \pi^2 m^2 L^2 w^{14}  \nonumber \\ &&
-12369505812480 a^6 \pi^3 m^6 w^{18}+61847529062400 a^4 \pi m^6 L^4 w^{18}-48241072668672 a^3 m^6 L^6 w^{18}  \nonumber \\ &&
+5098126180352 a^3 \pi^2 m^4 L^2 w^{16}-2796023709696 a^4 \pi^3 m^4 w^{16}-5497558138880 a^5 m^8 L^6 w^{20}  \nonumber \\ &&
+3208340570112 a^2 \pi m^4 L^4 w^{16}+1117765238784 a^2 \pi^3 m^2 w^{14}-65146063945728 a^5 \pi^2 m^6 L^2 w^{18}  \nonumber \\
r^{d}_{26}&:=& 27212912787456 a^4 \pi^3 m^4 w^{18}-370440929280 \pi^3 w^{14}+252887674388480 a^4 \pi m^6 L^4 w^{20}  \nonumber \\ &&
-39582418599936 a^5 \pi^2 m^6 L^2 w^{20}-481586092965888 a^3 m^6 L^6 w^{20}+6176162971648 a^2 \pi^3 m^2 w^{16}  \nonumber \\ &&
+8108898254848 a^2 \pi m^4 L^4 w^{18}-26800595927040 a^3 \pi^2 m^4 L^2 w^{18}+12446815223808 a \pi^2 m^2 L^2 w^{16}  \nonumber \\
r^{d}_{28}&:=& 422212465065984 a^4 \pi m^6 L^4 w^{22}+160941014515712 a \pi^2 m^2 L^2 w^{18}  \nonumber \\ &&
-2427721674129408 a^3 m^6 L^6 w^{22}-3208340570112 \pi^3 w^{16}+148434069749760 a^4 \pi^3 m^4 w^{20}  \nonumber \\ &&
-8383776161792 a^2 \pi^3 m^2 w^{18}-392525651116032 a^3 \pi^2 m^4 L^2 w^{20}  \nonumber \\ &&
-101155069755392 a^2 \pi m^4 L^4 w^{20}  \nonumber \\
r^{d}_{30}&:=& -949978046398464 a^2 \pi m^4 L^4 w^{22}-8108898254848 \pi^3 w^{18}  \nonumber \\ &&
+1242448139386880 a \pi^2 m^2 L^2 w^{20}-1266637395197952 a^3 \pi^2 m^4 L^2 w^{22}  \nonumber \\ &&
-270479860432896 a^2 \pi^3 m^2 w^{20}-5066549580791808 a^3 m^6 L^6 w^{24}  \nonumber \\
r^{d}_{32}&:=& 101155069755392 \pi^3 w^{20}+5383208929591296 a \pi^2 m^2 L^2 w^{22}   \nonumber \\ &&
-949978046398464 a^2 \pi^3 m^2 w^{22}-2533274790395904 a^2 \pi m^4 L^4 w^{24}    \nonumber \\
r^{d}_{34}&:=& 10133099161583616 a \pi^2 m^2 L^2 w^24+949978046398464 \pi^3 w^{22}   \nonumber \\
r^{d}_{36}&:=&2533274790395904 \pi^3 w^{24}
\end{eqnarray}
\end{scriptsize}
\section*{Appendix-II: Coefficients in Hawking Radiation Flux}
As stated earlier in the main section of the paper, the
thermodynamic metric tensor in the case of the fluctuation of the
energy flux is given by the Hessian matrix of the concerned energy
flux. In this case, the mass and the radiation frequency of the
Hawking radiating black hole configuration are respected to be the
intrinsic variables. Herewith, the energy flux fluctuations
indicate that the above two distinct parameters characterize the
intrinsic thermodynamic correlation functions. The present
computation shows that the exact set of correlations for both the
constant and decaying amplitude Hawking radiating black holes are
realized by employing the previously defined notations. Herewith,
we may easily enlist the coefficients of the thermodynamic
correlations arising from the fluctuations of the energy flux of a
Hawking radiating black hole. Both of the cases, \textit{viz,},
the constant amplitude and the decreasing amplitude configurations
demonstrate that the coefficients of the thermodynamic
correlations simplify to the following set of simple expressions.
\subsection*{Appendix (C): Constant Amplitude Flux Fluctuations}
In this appendix, we enlist explicitly expressions for the
constant amplitude thermodynamic correlations arising from the
thermodynamic fluctuations of the energy flux with an arbitrary
fluctuating mass and radiation frequency of a Hawking radiating
black hole. The functional nature of parameters, within a small
neighborhood of fluctuations introduced over an equilibrium
ensemble of configurations, may precisely be divulged as the
function of the radiation frequency and mass, with an exponential
form $\exp(8 \pi Mw)$. In the present framework, we can expose
that the coefficients of the constant amplitude Hawking radiating
black hole thermodynamic local and global correlations take the
following set of exact and extremely simple expressions.
\subsubsection*{Appendix (C1): Coefficients of the metric tensor}
\begin{scriptsize}
\begin{eqnarray}
c^{cMM}_0&:= & 3 \exp(24 \pi M w)-9 \exp(16 \pi M w)+9 \exp(8 \pi M w)-3 \nonumber \\
c^{cMM}_1&:=&4 u^2 \pi w \exp(16 \pi M w)-8 u^2 \pi w \exp(8 \pi M w)+4 u^2 \pi w \nonumber \\
c^{cMM}_2&:=&32 u^2 \pi^2 w^2 \exp(16 \pi M w)-32 u^2 \pi^2 w^2 \exp(8 \pi M w) \nonumber \\
c^{cMM}_3&:=&128 u^2 \pi^3 w^3 \exp(16 \pi M w)+128 u^2 \pi^3 w^3 \exp(8 \pi M w)  \nonumber \\
c^{cMw}_0&:=&\exp(16 \pi M w)+2 \exp(8 \pi M w)-1 \nonumber \\
c^{cMw}_1&:=&-8 \pi w \exp(16 \pi M w)+8 \pi w \exp(8 \pi M w) \nonumber \\
c^{cMw}_2&:=& 64 \pi^2 w^2 \exp(16 \pi M w)+64 \pi^2 w^2 \exp(8 \pi M w) \nonumber \\
c^{cww}_0&:=&\exp(8 \pi M w)+1)  \nonumber \\
c^{cww}_1&:=&4 \pi w \exp(8 \pi M w)+4 \pi w
\end{eqnarray}
\end{scriptsize}
\subsubsection*{Appendix (C2): Coefficients of the determinant of the metric tensor}
\begin{scriptsize}
\begin{eqnarray}
h^{c}_0&:=&-24 \exp(32 \pi M w)-3 u^2 \exp(8 \pi M w)-u^2 \exp(24 \pi M w)+u^2+24 \exp(8 \pi M w) \nonumber \\ &&
+3 u^2 \exp(16 \pi M w)+72 \exp(24 \pi M w)-72 \exp(16 \pi M w)  \nonumber \\
h^{c}_1&:=& -48 u^2 \pi w \exp(24 \pi M w)+96 \pi w \exp(8 \pi M w)+96 u^2 \pi w \exp(16 \pi M w) \nonumber \\&&
-96 \pi w \exp(16 \pi M w)-96 \exp(24 \pi M w) \pi w+96 \exp(32 \pi M w) \pi w-48 u^2 \pi w \exp(8 \pi M w) \nonumber \\
h^{c}_2&:=& -64 u^2 \pi^2 w^2 \exp(24 \pi M w)+320 u^2 \pi^2 w^2 \exp(16 \pi M w)-256 u^2 \pi^2 w^2 \exp(8 \pi M w) \nonumber \\
h^{c}_3&:=& 1024 u^2 \pi^3 w^3 \exp(16 \pi M w)+1024 u^2 \pi^3 w^3 \exp(24 \pi M w)
\end{eqnarray}
\end{scriptsize}
\subsubsection*{Appendix (C3): Coefficients of the scalar curvature}
\begin{scriptsize}
\begin{eqnarray}
l^{c}_0&:=& 9-u^2+81 \exp(40 \pi M w)-45 \exp(48 \pi M w)+9 \exp(56 \pi M w)+20 u^2 \exp(24 \pi M w)  \nonumber \\ &&
-45 \exp(8 \pi M w)-15 u^2 \exp(16 \pi M w)+6 u^2 \exp(40 \pi M w)-45 \exp(32 \pi M w)  \nonumber \\ &&
-\exp(48 \pi M w) u^2-45 \exp(24 \pi M w)+81 \exp(16 \pi M w)+6 u^2 \exp(8 \pi M w)  \nonumber \\ &&
-15 u^2 \exp(32 \pi M w)  \nonumber \\
l^{c}_1&:=& 24 \pi w-16 u^2 \pi w +888 \exp(48 \pi M w) \pi w+1680 \exp(24 \pi M w) \pi w  \nonumber \\ &&
+192 \pi w \exp(8 \pi M w)-240 \exp(56 \pi M w) \pi w-1080 \pi w \exp(16 \pi M w)  \nonumber \\ &&
+160 u^2 \pi w \exp(24 \pi M w)-160 u^2 \pi w \exp(16 \pi M w)-864 \exp(40 \pi M w) \pi w  \nonumber \\ &&
-80 u^2 \pi w \exp(32 \pi M w)+80 u^2 \pi w \exp(8 \pi M w)-600 \exp(32 \pi M w) \pi w  \nonumber \\ &&
+16 u^2 \pi w \exp(40 \pi M w)   \nonumber \\
l^{c}_2&:=& 1152 \pi^2 w^2 \exp(8 \pi M w)-448 u^2 \pi^2 w^2 \exp(24 \pi M w)-5760 \exp(40 \pi M w) \pi^2 w^2  \nonumber \\ &&
-32 u^2 \pi^2 w^2+11520 \exp(32 \pi M w) \pi^2 w^2+512 u^2 \pi^2 w^2 \exp(16 \pi M w)  \nonumber \\ &&
-288 u^2 \pi^2 w^2 \exp(32 \pi M w)-1152 \pi^2 w^2 \exp(48 \pi M w)-5760 \exp(24 \pi M w) \pi^2 w^2  \nonumber \\ &&
+1152 \exp(56 \pi M w) \pi^2 w^2+544 u^2 \pi^2 w^2 \exp(40 \pi M w)-1152 \pi^2 w^2 \exp(16 \pi M w)  \nonumber \\ &&
-192 u^2 \pi^2 w^2 \exp(48 \pi M w)-96 u^2 \pi^2 w^2 \exp(8 \pi M w)   \nonumber \\
l^{c}_3&:=& 1536 \pi^3 w^3 \exp(8 \pi M w)-3072 \exp(48 \pi M w) \pi^3 w^3+3072 \pi^3 w^3 \exp(16 \pi M w)  \nonumber \\ &&
-1536 \exp(56 \pi M w) \pi^3 w^3-8192 u^2 \pi^3 w^3 \exp(32 \pi M w)-10752 \pi^3 w^3 \exp(24 \pi M w)  \nonumber \\ &&
+2048 u^2 \pi^3 w^3 \exp(16 \pi M w)-1792 u^2 \pi^3 w^3 \exp(8 \pi M w)+3328 u^2 \exp(40 \pi M w) \pi^3 w^3  \nonumber \\ &&
+4608 u^2 \pi^3 w^3 \exp(24 \pi M w) +10752 \exp(40 \pi M w) \pi^3 w^3  \nonumber \\
l^{c}_4&:=&-6144 u^2 \pi^4 w^4 \exp(16 \pi M w)-28672 u^2 \pi^4 w^4 \exp(32 \pi M w)+4096 u^2 \pi^4 w^4 \exp(40 \pi M w) \nonumber \\ &&
-4096 u^2 \pi^4 w^4 \exp(8 \pi M w)+2048 u^2 \pi^4 w^4 \exp(48 \pi M w)+32768 u^2 \pi^4 w^4 \exp(24 \pi M w)  \nonumber \\
l^{c}_5&:=&-32768 \pi^5 w^5 \exp(40 \pi M w) u^2+32768 \pi^5 w^5 \exp(24 \pi M w) u^2
\end{eqnarray}
\end{scriptsize}
\subsection*{Appendix (D): Decaying Amplitude Horizon Fluctuations}
In the framework of the intrinsic geometry, we notice that we can
explicitly provide the exact expressions for the decaying
amplitude thermodynamic correlations arising from the fluctuations
of the energy flux with an arbitrary fluctuating mass of the
Hawking radiating black hole. Within a small neighborhood of
statistical fluctuations introduced over an equilibrium ensemble
of configurations, the functional nature of parameters may
precisely be computed as in the case of the constant amplitude
thermodynamic fluctuations of the energy flux. In this appendix,
we enlist the coefficients of the thermodynamic correlations
arising from the decaying amplitude energy flux of a Hawking
radiating black hole and thus show that they take the following
exact and simple expressions.
\subsubsection*{Appendix (D1): Coefficients of the metric tensor}
\begin{scriptsize}
\begin{eqnarray}
c^{dMM}_0&:=& 24 a^2 m^2 \pi w \exp(16 \pi M w)-48 a^2 m^2 \pi w \exp(8 \pi M w)+24 a^2 m^2 \pi w  \nonumber \\
c^{dMM}_1&:=& -3+3 \exp(24 \pi M w)-9 \exp(16 \pi M w)+9 \exp(8 \pi M w)  \nonumber \\ &&
+96 a^2 m^2 \pi^2 w^2 \exp(16 \pi M w)-96 a^2 m^2 \pi^2 w^2 \exp(8 \pi M w)  \nonumber \\
c^{dMM}_2&:=& 128 a^2 m^2 \pi^3 w^3 \exp(16 \pi M w)+128 a^2 m^2 \pi^3 w^3 \exp(8 \pi M w)  \nonumber \\
c^{dMw}_0&:=& 3 \exp(16 \pi M w)-6 \exp(8 \pi M w)+3  \nonumber \\
c^{dMw}_1&:=& 8 \pi w \exp(16 \pi M w)-8 \pi w \exp(8 \pi M w)  \nonumber \\
c^{dMw}_2&:=& 64 \pi^2 w^2 \exp(16 \pi M w)+64 \pi^2 w^2 \exp(8 \pi M w)  \nonumber \\
c^{dMw}_0&:=& 1-\exp(8 \pi M w) \nonumber \\
c^{dww}_1&:=& 4 \pi w \exp(8 \pi M w)+4 \pi w
\end{eqnarray}
\end{scriptsize}
\subsubsection*{Appendix (D2): Coefficients of the determinant of the metric tensor}
\begin{scriptsize}
\begin{eqnarray}
h^{d}_0&:= & -9 a^2 m^2 \exp(24 \pi M w)-27 a^2 m^2 \exp(8 \pi M w)  \nonumber \\ &&
+9 a^2 m^2+27 a^2 m^2 \exp(16 \pi M w))  \nonumber \\
h^{d}_1 &:=& 288 \exp(16 \pi M w) a^2 m^2 \pi w-144 \exp(24 \pi M w) a^2 m^2 \pi w  \nonumber \\ &&
-144 \exp(8 \pi M w) a^2 m^2 \pi w  \nonumber \\
h^{d}_2&:=& 832 \exp(16 \pi M w) a^2 m^2 \pi^2 w^2+72 \exp(24 \pi M w)-72 \exp(16 \pi M w)  \nonumber \\ &&
+320 \exp(24 \pi M w) a^2 m^2 \pi^2 w^2-1152 \exp(8 \pi M w) a^2 m^2 \pi^2 w^2  \nonumber \\ &&
-24 \exp(32 \pi M w)+24 \exp(8 \pi M w)  \nonumber \\
h^{d}_3&:=& 96 \exp(32 \pi M w) \pi w-96 \pi w \exp(16 \pi M w)+96 \pi w \exp(8 \pi M w) \nonumber \\ &&
+1024 \exp(16 \pi M w) a^2 m^2 \pi^3 w^3+1024 \exp(24 \pi M w) a^2 m^2 \pi^3 w^3  \nonumber \\ &&
-96 \exp(24 \pi M w) \pi w
\end{eqnarray}
\end{scriptsize}
\subsubsection*{Appendix (D3): Coefficients of the scalar curvature}
\begin{scriptsize}
\begin{eqnarray}
l^{d}_0&:=& 54 a^2 m^2 \exp(40 \pi M w)-9 a^2 m^2-9 a^2 m^2 \exp(48 \pi M w)+180 a^2 m^2 \exp(24 \pi M w) \nonumber \\ &&
-135 a^2 m^2 \exp(16 \pi M w)-135 a^2 m^2 \exp(32 \pi M w)+54 a^2 m^2 \exp(8 \pi M w) \nonumber \\
l^{d}_1&:=& -144 a^2 m^2 \pi w+1680 \exp(24 \pi M w) a^2 m^2 \pi w-1560 \exp(16 \pi M w) a^2 m^2 \pi w \nonumber \\ &&
-960 \exp(32 \pi M w) a^2 m^2 \pi w+744 \exp(8 \pi M w) a^2 m^2 \pi w+264 \exp(40 \pi M w) a^2 m^2 \pi w \nonumber \\ &&
-24 a^2 m^2 \exp(48 \pi M w) \pi w  \nonumber \\
l^{d}_2&:=& -288 a^2 m^2 \exp(48 \pi M w) \pi^2 w^2-180 \exp(24 \pi M w)-12 \exp(8 \pi M w) \nonumber \\ &&
+768 \exp(8 \pi M w) a^2 m^2 \pi^2 w^2+768 \exp(40 \pi M w) a^2 m^2 \pi^2 w^2 \nonumber \\ &&
-180 \exp(40 \pi M w)+72 \exp(16 \pi M w)-12 \exp(56 \pi M w)+72 \exp(48 \pi M w) \nonumber \\ &&
-480 \exp(32 \pi M w) a^2 m^2 \pi^2 w^2-288 a^2 m^2 \pi^2 w^2+240 \exp(32 \pi M w)\nonumber \\ &&
-480 \exp(16 \pi M w) a^2 m^2 \pi^2 w^2 \nonumber \\
l^{d}_3&:=& -1536 a^2 m^2 \pi^3 w^3 \exp(8 \pi M w)-21504 \exp(32 \pi M w) a^2 m^2 \pi^3 w^3 \nonumber \\ &&
+24 \exp(56 \pi M w) \pi w+96 \pi w \exp(16 \pi M w)-3584 \exp(16 \pi M w) a^2 m^2 \pi^3 w^3 \nonumber \\ &&
-120 \exp(24 \pi M w) \pi w+19456 \exp(24 \pi M w) a^2 m^2 \pi^3 w^3-96 \exp(48 \pi M w) \pi w \nonumber \\ &&
+120 \exp(40 \pi M w) \pi w+6656 \exp(40 \pi M w) a^2 m^2 \pi^3 w^3 \nonumber \\ &&
+512 a^2 m^2 \exp(48 \pi M w) \pi^3 w^3-24 \pi w \exp(8 \pi M w) \nonumber \\
l^{d}_4&:=& 3456 \exp(32 \pi M w) \pi^2 w^2-192 \exp(48 \pi M w) \pi^2 w^2 \nonumber \\ &&
-40960 \exp(32 \pi M w) a^2 m^2 \pi^4 w^4+65536 \exp(24 \pi M w) a^2 m^2 \pi^4 w^4 \nonumber \\ &&
+1024 \exp(40 \pi M w) a^2 m^2 \pi^4 w^4+288 \pi^2 w^2 \exp(8 \pi M w) \nonumber \\ &&
-1824 \exp(40 \pi M w) \pi^2 w^2+288 \exp(56 \pi M w) \pi^2 w^2 \nonumber \\ &&
+1024 \exp(48 \pi M w) a^2 m^2 \pi^4 w^4-17408 \exp(16 \pi M w) a^2 m^2 \pi^4 w^4 \nonumber \\ &&
-1824 \exp(24 \pi M w) \pi^2 w^2-9216 a^2 m^2 \pi^4 w^4 \exp(8 \pi M w) \nonumber \\ &&
-192 \exp(16 \pi M w) \pi^2 w^2  \nonumber \\
l^{d}_5&:=& 16384 \exp(24 \pi M w) \pi^5 w^5 a^2 m^2-768 \exp(56 \pi M w) \pi^3 w^3 \nonumber \\ &&
-16384 \exp(40 \pi M w) \pi^5 w^5 a^2 m^2+5376 \exp(40 \pi M w) \pi^3 w^3 \nonumber \\ &&
-1536 \exp(48 \pi M w) \pi^3 w^3+1536 \exp(16 \pi M w) \pi^3 w^3 \nonumber \\ &&
-5376 \exp(24 \pi M w) \pi^3 w^3+768 \pi^3 w^3 \exp(8 \pi M w)
\end{eqnarray}
\end{scriptsize}


\begin{thebibliography}{99}

\bibitem{Hawking} S. W. Hawking,
``Particle creation by black holes'',
Comm. Math. Phys. {\bf 43} 199 (1975).

\bibitem{semiclassical} N. D. Birrel and P. C. W. Davies,
``Quantum Fields in Curved Space'', Cambridge University Press, (1982).

\bibitem{4} C. R. Stephens, G. 't Hooft and B. F. Whiting,
``Black hole evaporation without information loss'',
Class. Quant. Grav. {\bf 11} 621 (1994).

\bibitem{5} T. Jacobson,
``Black hole radiation in the presence of a short distance
cutoff'', Phys. Rev. D {\bf 48} (1993) 728.

\bibitem{6} T. Jacobson, ``Black-hole evaporation and ultrashort
distances",  Phys. Rev. D {\bf 44} 1731 (1991).

\bibitem{7} Y. Kiem, H. Verlinde, E. Verlinde,
``Black hole complementarity versus locality'',
 Phys. Rev. D {\bf 52} 7053 (1995).

\bibitem{8} W. G. Unruh,
``Sonic analogue of black holes and the effects of high frequencies on black hole evaporation'',
 Phys. Rev. D {\bf 51} 2827 (1995).

\bibitem{11} P. Kraus and F. Wilczek,
``Self-interaction correction to black hole radiance'',
 Nucl. Phys. B {\bf 433} 403 (1995).

\bibitem{13} J. Schwinger,
``Brownian Motion of a Quantum Oscillator'', J. Math. Phys. {\bf 2} 407 (1961).

\bibitem{15} R. Feynman and F. Vernon,
``The theory of a general quantum system interacting with a linear dissipative system'',
Ann. Phys. (NY) {\bf 24} 118 (1963).

\bibitem{16} R. Feynman and A. Hibbs, ``Quantum Mechanics and Path Integrals'', McGraw-Hill, New York, (1965).

\bibitem{18} A. Campos and B. L. Hu,
``Fluctuations in a Thermal Field and Dissipation of a Black Hole Spacetime: Far-Field Limit'',
 Int. J. Theor. Phys. {\bf 38} 1253  (1999), {\tt http://arxiv.org/abs/gr-qc/9812034v1}.

\bibitem{19} R. Mart ́ and E. Verdaguer,
``On the semiclassical Einstein-Langevin equation'',
Phys. Lett. B {\bf 465} 113 (1999), {\tt http://arxiv.org/abs/gr-qc/9811070v2}.

\bibitem{FNB} V. Frolov and I. Novikov,
``Black Hole Physics: Basic Concepts and New Developments (Fundamental Theories of Physics)'',
Kluwer Academic Publ., ISBN-13: 978-0792351450, (1998).

\bibitem{BFP} C. Barrab\`{e}s , V. Frolov and R. Parentani,
``Metric Fluctuation Corrections to Hawking Radiation'',
Phys. Rev. D {\bf 59} 124010  (1999),
{\tt http://fr.arXiv.org/abs/gr-qc/9812076v2}.

\bibitem{York} J. W. York, Jr.,
``Dynamical origin of black-hole radiance'',
 Phys. Rev. D {\bf 28} 2929 (1983).

\bibitem{bntgup} B. N. Tiwari,
`` On Generalized Uncertainty Principle'', Proc. Quantum Gravity,
Hoelback 2008, {\tt arXiv:0801.3402v1 [hep-th]}.

\bibitem{sfm} S. Bellucci, S. Ferrara, A. Marrani,
``Attractors in Black'', Contribution to the Proceedings of the
3$^{rd}$ RTN Workshop Constituents, Fundamental Forces and
Symmetries of the Universe, 1-5 October 2007, Valencia, Spain,
Fortsch. Phys. {\bf 56} (2008) 761, {\tt arXiv:0805.1310}.

\bibitem{rup3} G. Ruppeiner,
``Thermodynamic curvature and phase transitions in Kerr-Newman black holes'',
Phys. Rev. D  {\bf 75} 024037 (2007), {\tt http://arxiv.org/abs/0802.1326v3}.

\bibitem{fgk} S. Ferrara, G. W. Gibbons, R. Kallosh,
``Black Holes and Critical Points in Moduli Space'',
Nucl. Phys. {\bf B500} 75 (1997), {\tt hep-th/9702103}.

\bibitem{cai1} J. Shen, R. G. Cai, B. Wang, R. K. Su,
``Thermodynamic Geometry and Critical Behavior of Black Holes'',
Int. J. Mod. Phys. A {\bf 22} 11 (2007), {\tt arXiv:gr-qc/0512035v1}.

\bibitem{gr-qc/0304015v1} J. E. Aman, I. Bengtsson, N. Pidokrajt,
``Geometry of black hole thermodynamics'',
Gen. Rel. Grav. {\bf 35} 1733 (2003), {\tt arXiv:gr-qc/0304015v1}.

\bibitem{0510139v3}
J. E. Aman, N. Pidokrajt,
``Geometry of Higher-Dimensional Black Hole Thermodynamics'',
Phys. Rev. D {\bf 73} 024017 (2006), {\tt arXiv:hep-th/0510139v3}.

\bibitem{Arcioni} G. Arcioni and E. Lozano-Tellechea,
``Stability and critical phenomena of black holes and black rings'',
Phys. Rev. D {\bf 72} 104021 (2005).

\bibitem{9601029v2} A. Strominger, C. Vafa,
``Microscopic Origin of the Bekenstein-Hawking Entropy,''
Phys. Lett. B {\bf 379} 99 (1996), {\tt arXiv:hep-th/9601029v2}.

\bibitem{9411187v3} A. Sen,
``Black Hole Solutions in Heterotic String Theory on a Torus,''
Nucl. Phys. B {\bf 440} 421 (1995), {\tt arXiv:hep-th/9411187v3}.

\bibitem{9504147v2} A. Sen,
``Extremal Black Holes and Elementary String States,''
Mod. Phys. Lett. A {\bf 10} 2081 (1995), {\tt arXiv:hep-th/9504147v2}.

\bibitem{0409148v2} A. Dabholkar,
``Exact Counting of Black Hole Microstates,'' Phys. Rev. Lett.
{\bf 94} (2005) 241301, {\tt arXiv:hep-th/0409148v2}.

\bibitem{9707203v1} L. Andrianopoli, R. D'Auria, S. Ferrara,
``Flat Symplectic Bundles of N-Extended Supergravities, Central Charges and Black-Hole Entropy,''
{\tt arXiv:hep-th/9707203v1}.

\bibitem{0507014v1} A. Dabholkar, F. Denef, G. W. Moore, B. Pioline,
``Precision Counting of Small Black Holes,''
JHEP  {\bf 096} 0510 (2005), {\tt arXiv:hep-th/0507014v1}.

\bibitem{0502157v4} A. Dabholkar, F. Denef, G. W. Moore, B. Pioline,
``Exact and Asymptotic Degeneracies of Small Black Holes,'' JHEP
{\bf 021} 0508 (2005), {\tt arXiv:hep-th/0502157v4}.

\bibitem{0505122v2} A. Sen,
``Stretching the Horizon of a Higher Dimensional Small Black Hole,''
JHEP {\bf 073} 0507 (2005), {\tt arXiv:hep-th/0505122v2}.

\bibitem{0209114} J. P. Gauntlett, J. B. Gutowski, C. M. Hull, S. Pakis, H. S. Reall,
``All supersymmetric solutions of minimal supergravity in five dimensions,''
Class. Quant. Grav. {\bf 20} 4587  (2003), {\tt arXiv:hep-th/0209114v3}.

\bibitem{0401129} J. B. Gutowski, H. S. Reall,
``General supersymmetric AdS5 black holes,'' JHEP {\bf 048} 0404
(2004), {\tt arXiv:hep-th/0401129v3}.

\bibitem{0602015v3}
L. Bonora, C. Maccaferri, R. J. Scherer Santos, D. D. Tolla,
``Bubbling AdS and Vacuum String Field Theory," Nucl. Phys. B {\bf
749} 338 (2006), {\tt arXiv:hep-th/ 0602015v3}.

\bibitem{0408106} I. Bena, N. P. Warner,
``One Ring to Rule Them All ... and in the Darkness Bind Them ?,''
Adv. Theor. Math. Phys. {\bf 9} 667 (2005), {\tt arXiv:hep-th/0408106v2}.

\bibitem{0408122} J. P. Gauntlett, J. B. Gutowski,
``General Concentric Black Rings,''
Phys. Rev. D {\bf 71} 045002 (2005), {\tt arXiv:hep-th/0408122v3}.

\bibitem{SJJS} M. M. Sheikh-Jabbari, J. Simon,
``On Half-BPS States of the ABJM Theory," JHEP {\bf 073} 0908
(2009), {\tt arXiv:0904.4605v2 [hep-th]}.

\bibitem{BBJS} V. Balasubramanian, J. de Boer, V. Jejjala, J. Simon
`` Entropy of near-extremal black holes in $AdS_5$,"
 JHEP {\bf 067} 0805 (2008), {\tt arXiv:0707.3601}.

\bibitem{GLS}  E. G. Gimon, F. Larsen, J. Simon
``Black Holes in Supergravity: the non-BPS Branch," JHEP {\bf 040}
0801 (2008), {\tt arXiv:0710.4967v3 [hep-th]}.

\bibitem{0606084v1} T. Sarkar, G. Sengupta, B. N. Tiwari,
``On the Thermodynamic Geometry of BTZ Black Holes,'' JHEP {\bf
0611} (2006) 015, {\tt arXiv:hep-th/0606084v1}.

\bibitem{SST} T. Sarkar, G. Sengupta, B. N. Tiwari,
``Thermodynamic Geometry and Extremal Black Holes in String
Theory,'' JHEP {\bf 0810} 076 2008, {\tt arXiv:0806.3513v1
[hep-th]}.

\bibitem{bnt} B. N. Tiwari,
``Sur les corrections de la g\'{e}om\'{e}trie thermodynamique des trous noirs,''
{\tt arXiv:0801.4087v1 [hep-th]}.

\bibitem{BNTBull} S. Bellucci, B. N. Tiwari,
``On the Microscopic Perspective of Black Branes Thermodynamic Geometry,''
 Proc. Quantum Gravity, Hoelback 2008, {\tt arXiv:0808.3921v1 [hep-th]}.

\bibitem{SAMpaper} S. Bellucci, B. N. Tiwari,
``An Exact Fluctuating 1/2-BPS Configuration'', JHEP {\bf 05} 023
(2010), {\tt arXiv:0910.5314v1 [hep-th]}.

\bibitem{bntSb} S. Bellucci, B. N. Tiwari,
``State-space Manifold and Rotating Black Holes'', {\tt To appear}.

\bibitem{BSBR} S. Bellucci, B. N. Tiwari,
``Black Strings, Black Rings and State-space Manifold'', {\tt To appear}.

\bibitem{RuppeinerPRD78} G. Ruppeiner,
``Thermodynamic curvature and phase transitions in Kerr-Newman
black holes,'' Phy. Rev. D {\bf 78} 024016 (2008).

\bibitem{RuppeinerRMP} G. Ruppeiner,
``Riemannian geometry in thermodynamic fluctuation theory,'' Rev.
Mod. Phys {\bf 67} 605 (1995), Erratum  {\bf 68} (1996) 313.

\bibitem{RuppeinerA20} G. Ruppeiner,
``Thermodynamics: A Riemannian geometric model,'' Phys. Rev. A
{\bf 20} 1608 (1979).

\bibitem{BonoraCvitan} L. Bonora, M. Cvitan,
``Hawking radiation, W-infinity algebra and trace anomalies", JHEP
 {\bf 0805} 071 (2008), {\tt arXiv:0804.0198v3 [hep-th]}.

\bibitem{BSKS1} S. Bellucci, S. Krivonos, A. Sorin,
``The $W(sl(N+3), sl(3))$ algebras and their contractions to
$W_3$", Phys. Lett. B {\bf 392} 350 (1997), {\tt arXiv:
hep-th/9609038}.

\bibitem{BSKS2} S. Bellucci, S. Krivonos, A. Sorin,
``Null Fields Realizations of $W_3$ from $W(sl(4),sl(3))$ and
$W(sl(3|1),sl(3))$ Algebras", Phys. Lett. B {\bf 366} 104 (1996),
{\tt arXiv: hep-th/9509072}.

\bibitem{BSKS3} S. Bellucci, S. Krivonos, A. Sorin, ``Linearizing $W_{2,4}$ and $WB_2$
Algebras," Phys. Lett. B {\bf 347} 260 (1995), {\tt arXiv:
hep-th/9411168}.

\bibitem{BSKS4} S. Bellucci, S. Krivonos, A. Sorin,
``Nonlinear Realizations of the $W_3^{(2)}$ Algebra," Phys. Lett.
A {\bf 191} 216 (1994), {\tt arXiv:hep-th/9402151}.

\bibitem{RuppeinerPRL} G. Ruppeiner,
``Thermodynamic Critical Fluctuation Theory?,'' Phys. Rev. Lett.
{\bf 50} 287 (1983).

\bibitem{RuppeinerA27} G. Ruppeiner,
``New thermodynamic fluctuation theory using path integrals,''
Phys. Rev. A {\bf 27} 1116 (1983).

\bibitem{RuppeinerA41} G. Ruppeiner and C. Davis,
``Thermodynamic curvature of the multicomponent ideal gas,'' Phys.
Rev. A {\bf 41} 2200 (1990).

\bibitem{Bekenstein} D. Bekenstein,
``Information in the holographic universe'', Sci. Am. {\bf 289},
No. 2, 58-65 (2003).

\bibitem{BNTSBVC} S. Bellucci, V. Chandra, B. N. Tiwari,
``On the Thermodynamic Geometry of Hot QCD,''
{\tt arXiv:0812.3792v1 [hep-th]}.

\bibitem{ZCZ} B. Zhang, Q. Y. Cai, M. S. Zhan
``Hawking radiation as tunneling derived from Black Hole:
Thermodynamics through the quantum horizon'',
Phys. Lett. B {\bf 665} 260 (2008),
{\tt http://arXiv.org/abs/0806.2015v1}.

\end{thebibliography}
\end{document}